\documentclass[twocolumn,showpacs,pre,superscriptaddress]{revtex4}
\usepackage{amsmath,amsfonts,amssymb}
\usepackage{graphicx}

\newcommand{\be}{\begin{equation}} 
\newcommand{\ee}{\end{equation}} 
\newcommand{\bea}{\begin{eqnarray}} 
\newcommand{\eea}{\end{eqnarray}}  
\newcommand{\bean}{\begin{eqnarray*}} 
\newcommand{\eean}{\end{eqnarray*}}

\def\lsim{\raise 0.4ex\hbox{$<$}\kern -0.8em\lower 0.62ex\hbox{$\sim$}} 
\def\gsim{\raise 0.4ex\hbox{$>$}\kern -0.7em\lower 0.62ex\hbox{$\sim$}}

\begin{document}

\title{A method of generating initial
conditions for cosmological N body simulations}

\author{M. Joyce} \affiliation{Laboratoire de Physique
Nucl\'eaire et de Hautes Energies, Universit\'e de Paris VI,\\ 4,
Place Jussieu, Tour 33 -RdC, 75252 Paris Cedex 05, France.}

\author{D. Levesque} \affiliation{Laboratoire de Physique
Th\'eorique, Universit\'e de Paris XI, B\^atiment 210, 91405 Orsay,
France.}

\author{B. Marcos} \affiliation{Laboratoire de Physique
Th\'eorique, Universit\'e de Paris XI, B\^atiment 210, 91405 Orsay,
France.}

\begin {abstract}
We investigate the possibility
of generating initial conditions for cosmological N-body 
simulations by simulating a system whose correlations at 
thermal equilibrium approximate well those
of cosmological density perturbations. The system is 
an appropriately modified version of the standard 
``one component plasma'' (OCP). We show first how a well-known 
semi-analytic method can be used to determine the potential 
required to produce the desired correlations, and then verify 
our results for some cosmological type spectra with simulations 
of the full molecular dynamics. The advantage of the method, compared 
to the standard one, is that it gives by construction an
accurate representation of both the real and reciprocal
space correlation properties of the theoretical model.
Furthermore the distributions are also statistically homogeneous
and isotropic. We discuss briefly the modifications needed to implement 
the method to produce configurations appropriate for large 
N-body simulations in cosmology, and also the generation of 
initial velocities in this context.

\end{abstract}

\maketitle

\section{INTRODUCTION}
One of the major goals of current models in cosmology is
to link the very small amplitude fluctuations observed in the 
cosmic microwave background to the large inhomogeneities
observed today in the distribution of galaxies. The former
should serve as the initial seeds for the formation of the
large structures observed, in the available time, through the 
attraction of gravity. In the absence of a full analytic 
understanding of the problem, gravitational N-body simulation
has become an essential instrument in making this link
(for reviews, see e.g. \cite{EDFW}, \cite{white}, \cite{couchman}). 
In this paper we are concerned with one element of N-body
simulation: the setting up of appropriate initial conditions
(IC). 

The theoretical IC for such simulations are specified by the cosmological
model considered, which gives the correlation properties, at the initial time
of the simulation, of a {\it continuous} stochastic mass density field
$\rho(\vec{r})$.  This field has a well defined mean density $\rho_0 (>0)$,
and fluctuations $\delta_{\rho}(\vec{r})= \rho(\vec{r})/\rho_0 -1$, which are
of small amplitude an generically assumed to be Gaussian. The correlation
properties are thus fully specified by the power spectrum (PS) of the model.
The subtlety in the generation of the IC for an N-body simulation is in the
fact that a realization of this theoretical stochastic density field, which is
a {\it continuous} function, must be discretised: it must be represented by an
initial distribution in space of the N point particles used in the simulation.
The correlation properties of these two qualitatively different kinds of
distributions, however, cannot be the same.  At small scales, in particular,
the discrete distribution has density fluctuations with infinite variance
(points), while the continuous model has finite variance fluctuations at all
scales. The correlation properties of the set of N discrete particle must,
nevertheless, approximate, in some appropriate sense, those of the theoretical
model.

There is one well known and universally employed method for generating the IC
of N body simulations. It uses the so-called ``Zeldovich approximation''
\cite{zel, buchert}.  Particles are displaced off a perfect lattice (or
sometimes ``glassy'') configuration in a manner determined by the desired
theoretical PS. Thus every realization of the theoretical stochastic density
field is mapped onto an initial configuration of the N particles. Each such
configuration can thus be considered as a realizaion of a {\it discrete}
stochastic density field.

Usually the characterisation of the correlation properties of this discrete
distribution (i.e. of the generated discrete IC) are considered only in
reciprocal space, i.e. in terms of a measured PS, which approximates well the
PS of the input theoretical model for sufficiently small $k$ (i.e. small
compared to the Nyquist frequency characteristic of the discreteness of the
underlying lattice). The question of the representation of real space
correlation properties (e.g. mass variance in spheres and the two point
correlation function) has been the subject of some discussion in the recent
literature \footnote{In \cite{slb} direct measures of the mass variance in
  spheres, and the two point correlation function, from an initial
  configuration for a large N-body simulation of the VIRGO consortium
  \cite{jenkins98}, are reported. The authors conclude that these quantities
  do not agree at all with their corresponding theoretical values. In
  \cite{kdcomment} the same analysis was redone, and an error in the
  normalisation in \cite{slb} of the theoretical variance was identified. From
  their comparison the authors then concluded that the agreement between
  measured and theoretical properties was good for the variance, but that no
  conclusion could be drawn about the two point correlation function, which
  they argue was dominated by statistical noise. In a further article \cite
  {kd} the same authors analyse these quantities in the IC of a set of
  simulations, and arrive at the same conclusions i.e.  agreement for the
  variance, and no conclusion possible concerning the correlation function. We
  believe is it fair to say, however, that what is considered here as
  ``agreement'' of the theoretical and measured variance is somewhat
  subjective. The interested reader can evaluate the answer for himself by
  consulting the figures in \cite{kdcomment}, and Fig. 3 and Fig. 6 of \cite
  {kd}.}.  A detailed study of this question is presented in a recent article
by two of us \cite{jm04}. The standard method is shown rigorously to give a
distribution with a PS approximating well that of the input theoretical model
below the Nyquist frequency of the lattice, for quite a broad class of input
PS \footnote{The class of input spectra in three spatial dimensions for which
  the algorithm gives an accurate representation in $k$ space are those with a
  behaviour $P(k) \sim k^n$ at small $k$, with $-1<n\leq 2$.  For $n\leq -1$
  the amplitude of the displacement field (which is simply proportional to the
  gravitational acceleration) is not well defined (i.e. becomes a function of
  the size of the periodic box). For an input spectrum with $n \geq 2$ at
  small $k$, the generated distribution always has a power spectrum with $n=2$
  at small $k$.}.  In real space, however, it can be shown that the
correlation properties are generically a mixture, at all scales, between those
of the underlying lattice and those of the theoretical model.  The real space
correlation function carries, notably, the trace of the underlying oscillating
behaviour of the initial lattice correlation function. The mass variance in a
sphere is also typically dominated by that of the lattice to scales which,
depending on the input theoretical model and its normalisation, can be very
much larger than the interparticle spacing.

The question which inevitably follows is whether such a discrepancy between
the correlation properties of the theoretical model and those of its
discretisation are important for what concerns the interpretation of the
results of the dynamical evolution of the discrete N-body simulation under
gravity. These evolved configurations are taken to represent, over a very wide
range of scale (typically from well below the initial inter-particle
separation up to scales a little smaller than the size of the periodic box),
the evolution of a continuous self-gravitating fluid with the initial
theoretical correlation properties. Is it sufficient to have only a faithful
representation of the initial correlation properties in reciprocal space? This
is an open question, which we do not attempt to answer here. In order to
answer it, however, it would evidently be useful to have other methods for
generating discrete representations of the theoretical IC. More generally, for
the study of gravitational evolution from near homogeneous IC, it is of
interest to have a diversity of methods for generating initial configurations.

The aim of this work is to explore a possible such method which has been
suggested in \cite{gjjlpsl02}.  The idea is simple: to find a system whose
dynamics produces configurations with the required correlation properties.
Given that the IC describe very small fluctuations about homogeneity, it is
natural to seek a system whose thermal equilibrium gives rise to such
configurations. In \cite{gjjlpsl02} it has been noted that a simple
modification of the interparticle potential in the ``one component plasma''
(OCP), which is a well-known and much studied system in statistical physics,
gives rise to a PS at small wavenumbers which is identical to that of
cosmological models, with fluctuations which are also Gaussian. Further it is
argued qualitatively in \cite{gjjlpsl02} that an appropriate modification of
this potential at smaller scales could give rise to the correlations typical
of standard ``cold dark matter'' (CDM) cosmologies (which are of the class
currently favoured by cosmologists). In this paper we show how a well known
semi-analytic method, the hypernetted chain (HNC), for determining the
correlation properties of such systems at equilibrium can be adapted to allow
one to infer the required interaction potential given an input spectrum. We
apply this inversion for some simple model CDM-type spectra, and find a
potential with the qualitative features anticipated by \cite{gjjlpsl02}. We
then verify the accuracy of this method with a simulation of the full
molecular dynamics. This involves modifying the Ewald sum method for
calculation of the potential.

The paper is organised as follows. In the next section we discuss first the
important underlying question of how the correlation properties of a
continuous field may be represented by a discrete distribution. In section
\ref{hnc} we briefly recall the basic features of the standard OCP, and
explain the analytical statistical mechanics approach to the determination of
its correlation properties.  In particular we describe the HNC approximation
which gives a closed form for the determination of the two point correlation
properties from the potential.  We explain how it can be used in an inverted
form to infer the required inter-particle potential for an input (physically
reasonable) correlation function.  We show then how this can be done for a CDM
type spectra, giving numerical results for some cases.  In the next section we
turn to the simulation of the molecular dynamics of these systems. We give
results for several simulations which verify the results anticipated from the
HNC, and provide actual realizations of discretisations of CDM type spectra.
In the final section we discuss the possibility of using this method to
generate configurations representing the IC of cosmological NBS, as in
principle the method appears to work well and to provide an interesting, and
perhaps more satisfactory, alternative to the standard algorithm. One issue is
the generation of appropriate initial velocities. We explain that this can be
done as in the standard method, as the Zeldovich approximation gives a
relation between the initial particle velocity and the gravitational field
acting on it, which can also be applied to our configurations.  A more
non-trivial issue is that of the size of the configurations generated: the
simulations we have performed to test the viability of the method are of a
quite modest size --- the largest simulation reported here has slightly less
than 9000 particles --- while current cosmological simulations use, at the
very least, hundreds of thousands of particles.  Just as in the case of NBS
itself, it will be necessary to speed up the algorithm developed here to
attain this goal. The methods for this latter step already exist and can be
applied to the more complicated two body potentials we study.

\section{REPRESENTATION OF CONTINUOUS SPECTRA WITH POINT DISTRIBUTIONS}
\label{discretisation}

\subsection{Discrete and continuous stochastic density fields}

Let us first recall (see e.g. \cite{pee80}, \cite{gsljp-springer}) some basic
properties of the PS (or, in the terminology more common in statistical
physics, structure function).  We will use $P(\vec{k})$ to denote this
quantity when we refer to a continuous distribution, $S(\vec{k})$ for the
discrete case. For both cases it can be defined\footnote{We use here the
  convention from the cosmological literature in which the PS usually has
  dimensions of volume. Our definition of $S(k)$ then differs from the one
  standard in statistical physics, which makes $S(k)$ dimensionless, by a
  factor of the mean particle density $n_0$.} as
\begin{equation}
P(\vec{k})= \lim_{V\rightarrow \infty} 
\frac{\langle |\delta_{\rho}(\vec{k})|^2 \rangle}{V} 
\end{equation}
where $\delta(\vec{k})$ is the Fourier Transform (hereafter FT) of the
normalized fluctuation field $(\rho(\vec{r})-\rho_0)/\rho_0$, and the angle
brackets indicate an ensemble average. We will assume that the system is
statistically isotropic so the PS will not depend on the orientation of the
vector $\vec{k}$ i.e. $P(\vec{k})=P(k)$.  We will also assume \footnote{In
  assuming statistical homogeneity and isotropy we exclude formally the
  standard case of a perturbed lattice, which is not in this class. The
  results which are quoted below for that case are, nevertheless, valid (see
  \cite{jm04,gab04}).}  statistical homogeneity (i.e. invariance of average
quantities under translation). In this case the Fourier transform of the PS,
for which we use the convention
\begin{equation}
\xi(\vec r)=\frac{1}{(2\pi)^3}
\int d^3 k e^{i{\vec k}{\vec{r}}} P(\vec k)\,,
\label{FT}
\end{equation}
is the reduced two point correlation function defined as \be \xi(\vec{r})
\equiv \xi(r)= \frac{\langle \rho(r)\rho(0) \rangle}{\langle \rho (0)
  \rangle^2} -1 =\langle \delta_{\rho}(r) \delta_{\rho}(0) \rangle \,.
\label{red-2point}
\ee The intrinsic difference between a continuous and discrete density field
$\rho(\vec{r})$ manifests itself in a qualitative difference between the
mathematical properties of the two-point quantities in each case. In real
space the correlation function $\xi(\vec{r})$ has, for the class of finite
one-point variance continuous fields which we consider, the property \be -1
\leq \xi(r) \leq \xi(0) < \infty\,.
\label{props-cns-2pt}
\ee For the discrete case the one-point variance, which is equal to $\xi(0)$,
necessarily diverges because of the singular nature of the density field at
any point. The correlation function can then be written
\begin{equation}
\label{discrete_xi}
\xi(r)=\frac{1}{n_0} \delta (\vec{r}) + h(r)
\end{equation}
where $n_0$ is the mean number density, $\delta(\vec{r})$
is the (three dimensional) Dirac delta function, and $h(r)$
is  a non-singular function for all $r$ which
can be taken to have the property analogous to 
Eq.(\ref{props-cns-2pt}). 

These properties in
real space translate in $k$ space into
a difference in the asymptotic properties
of the power-spectra at large $k$. The one-point 
variance of the density  field is also given 
by the integral of the PS, and so for the 
continuous case
we have
\begin{equation}
\lim_{k \rightarrow \infty} k^3P(k) =0
\label{conPS-largek}
\end{equation}
in order that this variance be finite.
In the discrete case, on the other hand,
we have
\begin{eqnarray} 
\label{limit-largek}
\lim_{k \rightarrow \infty} S(k)= \frac{1}{n_0} \\
\lim_{k \rightarrow \infty} k^3(S(k) -\frac{1}{n_0}) =0 \,. 
\label{asym-largek}
\end{eqnarray}
i.e. the divergence of the one-point variance is entirely associated to the
``Poissonian'' term in the PS, which is simply the FT of the delta-function
singularity in real space explicit in Eq.(\ref{discrete_xi}). Note that both
$P(k)$ and $S(k)$ are, by definition, positive functions, while
$S(k)-\frac{1}{n_0}$ is not. There is therefore no bound $S(k) \geq 1/n_0$. In
particular, one can have $S(k) \rightarrow 0$ for $k \rightarrow 0$, in
systems satisfying the constraint \be \int d^3r h(r)=-\frac{1}{n_0}\, \ee i.e.
when there is appropriate anti-correlation to balance the contribution to
fluctuations at all scales from the Poissonian term associated to any discrete
process. As discussed in \cite{gjjlpsl02}, \cite{gsljp-springer}, \cite{glass}
these correspond to highly ordered ``super-homogeneous'' systems.

\subsection{Smoothing of discrete distributions}
\label{Smoothing of discrete distributions}

The intuitively evident fact that a discrete distribution can only represent
the correlation properties of a continuous field above some scale --- that
characteristic of the ``granularity'' of the discrete distribution --- is
reflected mathematically in the differences just discussed between the
properties in the two cases of the correlation function at small real space
separations, and the PS at large wave-numbers. Let us suppose now that we have
a discrete distribution with PS $S(k)$, and a continuous distribution with PS
$P(k)$.  What is meant when one says that the former is a discretisation of
the latter? In what sense can we say that the former represents the
correlation properties of the continuous distribution with PS $P(k)$?  The
answer to this question is that there is in fact no unique prescription for
passing between a discrete and continuous distribution. In particular taking
formally the limit in which the number of particles goes to infinity at fixed
mass density, which one might naively think to define the desired continuous
limit, does not do so. Consider, for example, the case of an (uncorrelated)
Poisson point process: as the number density is taken to infinity the
fluctuations also go to zero. Thus the continuous limit is an exactly uniform
distribution with $P(k)=0$.

As discussed in \cite{gjjlpsl02} the most 
natural way of defining such a relationship is by an 
appropriate local smoothing i.e. we assume the represented 
density field is given by the convolution of the discrete 
distribution with 
a spatial window function $W_{R_s}(r)$
\begin{equation}
\rho_c(\vec{r})= \int W_{R_s}(|\vec{r}-\vec{r}\,'|) \rho_d(\vec{r}\,') d^3 \vec{r}\,'
\end{equation}
where $R_s$ is the (single) characteristic smoothing scale and the
realization of the discrete field is a sum over all the particles
\be
\rho_d(\vec r)=\sum_i\delta(\vec r-\vec r_i) \,,  
\ee 
and $\rho_c(\vec{r})$ is the corresponding realization of the 
continuous stochastic density field.
We then have that
\begin{equation}
P(k) = |\tilde{W}_{R_s}(k)|^2 S(k)
\label{smoothing}
\end{equation}
where $\tilde{W}_{R_s}(k)$ is the Fourier transform of $W_{R_s}(r)$. By the
assumption that the window function gives a local smoothing, we mean that it
is an integrable function. It is naturally normalised to unity (to conserve
mass) so that $\tilde{W}_{R_s}(0)$ is equal to unity. Thus the PS of the
discrete field must approximate well that of the continuous one for small $k$
(i.e. $k \ll R_s^{-1}$).  In real space the smoothing leads to the convolution
relation 
\be \xi_c (r) = \int W_{R_s}(\vec r\,') W_{R_s}(\vec r\,'')
\xi_d(\vec r+\vec r\,'-\vec r\,'') d^3\vec r\,' d^3\vec r\,''
\label{smoothed-corr-fn}
\ee
between the continuous correlation function $\xi_c(r)$
and the discrete correlation function $\xi_d (r)$.
One sees explicitly how the singularity becomes
regularised applying (\ref{smoothed-corr-fn}) to (\ref{discrete_xi}):
\be
\frac{1}{n_0}\delta(\vec r)\to\frac{1}{n_0}\int W_{R_s}(\vec r\,') W_{R_s}(\vec r+\vec r\,')d^3\vec r\,'.
\ee

Note that any pair consisting of a discrete and a continuous density field,
with PS $P(k)$ and $S(k)$ respectively, can be related to one another formally
by Eq.(\ref{smoothing}), taken simply as a definition of the smoothing
function \footnote{This is evidently actually a family of functions as one has
  the freedom to choose an arbitrary phase factor as a function of $k$ when
  inverting the expression to obtain a $W_{R_s}(r)$.}. Whether $S(k)$ can be
considered to be a physically reasonable discretisation of $P(k)$ depends then
on the mathematical properties of this smoothing function i.e. whether it
really represents a physical smoothing.  It is useful, for what follows, to
express the relation between the two spectra in a slightly different (but
equivalent) form:
\begin{equation}
S(k)= P(k) + \frac{1}{n_0}D(k)
\label{prescription}
\end{equation}
where $n_0$ is the number density of the discrete 
distribution, The function $D(k)$ has then the 
properties imposed by Eqs.(\ref{limit-largek})
and (\ref{asym-largek}):
\begin{eqnarray}
\label{asym-D-1}
\lim_{k \rightarrow \infty} D(k)= 1 \\
\lim_{k \rightarrow \infty} k^3(D(k) -1)=0 
\,. 
\label{asym-D-2}
\end{eqnarray}
In real space one has analogously 
\begin{equation}
h(r)=\xi_c(r) - \frac{1}{n_0} FT [1-D(k)]
\label{corrfn-condition2}
\end{equation}
where $\xi_c(r)$ is the Fourier transform of
$P(k)$ i.e. the reduced two-point correlation 
function of the continuous model.
Expressed in terms of the smoothing 
we have from Eq.(\ref{smoothing}) 
that 
\begin{equation}
|\tilde{W}_{R_s}(k)|^{-2}=1 + \frac{D(k)}{n_0 P(k)} \,.
\label{smoothing-D}
\end{equation}
Note that whether the smoothing which is associated
to a $D(k)$ is a physical smoothing depends, therefore,
not only on its own properties, but also on those of 
$P(k)$. 

\subsection{Discretisation in the standard algorithm}

The standard method for generating IC for N-body simulations in cosmology
superimposes correlated displacements on a ``pre-initial'' configuration of
points representing ``uniformity''. The latter is usually taken to be a
perfect lattice, but sometimes also a ``glassy'' configuration obtained by
running the N-body code with the sign of gravity reversed (i.e. with Coulomb
forces) and an appropriate damping term. Both of these configurations
represent an unstable equilibrium point (approximately in the latter case)
under their self-gravity, and evolve only when the perturbations are applied.
The precise sense in which this method leads to a representation of the
continuous model (and whether this is adequate for the dynamics which is
simulated) is not discussed in the literature, but implicitly the criterion
used appears to be \be
\label{criterion-standard}
S(k) \approx P(k) \quad {\rm for} \quad k < k_d
\ee
where $k_d$ is the wave-number characteristic of the 
discreteness (i.e. the Nyquist frequency in the case
of the lattice). 
In \cite{jm04} we have 
investigated systematically the real and reciprocal space 
correlation properties of configurations generated
by this algorithm, using a formalism developed in \cite{gab04}. We give, in
particular, exact expressions for the PS $S(k)$ of the generated
distribution, for a given input theoretical PS $P(k)$. From this 
one can infer explicitly the form of
$D(k)$, as defined in Eq.(\ref{prescription}),  and 
the corresponding ``smoothing'' given
through Eq.(\ref{smoothing-D}). The essential point
is that $D(k)$ carries all the correlation of the 
``pre-initial'' configuration. Indeed one has, to 
a first approximation, simply that $D(k) \approx n_0 P_{in}(k)$ 
where  $P_{in}(k)$ is the PS of the ``pre-initial'' 
(lattice or glass) configuration.In $k$-space 
this correlation is ``localised'' at large $k$,
so that Eq.(\ref{criterion-standard}) is satisfied.
In real space, however, the second term on the right
hand side of Eq.~(\ref{corrfn-condition2} is not
localised to below the inter-particle distance,
and completetly dominates the
small amplitude correlations of the input model
\footnote{We neglect here the complication
introduced by the fact that lattice is not statistically 
isotropic. See \cite{jm04,gab04} for further detail.}.
Correspondingly the algorithm generically does not give a
discretisation which can be interpreted as a localised
smoothing in the sense we have discussed above:
the function $W_{R_S}(r)$ determined through
Eq.(\ref{smoothing}) 
(or, equivalently, Eq.(\ref{smoothing-D})
is also highly delocalised). This is simply because
the underlying
regularity of the ``pre-initial'' configuration, which
is not a feature of the continuous model, cannot
be removed by a localised smoothing.

\subsection{Determination of the PS of a new discretisation}
\label{Determination of the PS of a new discretisation}

We investigate here a different method for discretising 
a given input PS $P(k)$. The principle is to seek to generate 
a distribution with an $S(k)$ given through 
Eq.(\ref{prescription}), where for $D(k)$ we will choose
a smooth function of $k$, characterised by a single
scale $k_d$, and interpolating between
zero for $k < k_d$ and unity for $k > k_d$ (and
in keeping with the asymptotic properties
required Eqs.(\ref{asym-D-1}) and (\ref{asym-D-2})).
The scale $k_d$ will be chosen of order the
inverse of the mean particle separation $a$ (see 
below for the exact definition we use).
Further the function $D(k)$ 
will be such that the
FT of $(D(k)-1)$ in Eq.(\ref{corrfn-condition2})  
is localised strongly in real space on the 
scale $a$. Thus, by construction, we will converge to
\begin{eqnarray}
\label{convergence1}
S(k) &\approx& P(k) \quad {\rm for} \quad k \ll  k_d \\
h(r) &\approx& \xi_c (r) \quad {\rm for} \quad r \gg a
\label{convergence2}
\end{eqnarray}
As we have noted, whether this choice of $D(k)$ 
corresponds to a physical smoothing, in the sense
we have discussed, depends also on the properties
of $P(k)$. For the well-behaved $P(k)$ we will
consider we expect this to be the case, but we
will check explicitly that the function $W_{R_s}(r)$ 
is smooth and integrable.

The precise scale $k < k_d$ at which Eq.(\ref{convergence1}) holds will depend
both on $D(k)$ and on the form and normalisation of the PS.  In the
cosmological context $P(k)$ is generically a monotonically decreasing function
over a wide range of $k$ for $k < k_d$, and thus the dimensionless quantity
$n_0P(k_d)$ gives a parametrisation of the relative amplitudes of the
``continuous'' and ``discrete'' parts of the full PS $S(k)$.  In the
simulations of molecular dynamics described below we will take $n_0P(k_d) \sim
1$.  Thus we will have in this case Eq.(\ref{convergence1}) for all $k \lsim
k_d$, and (we will verify) Eq.(\ref{convergence2}) from $r \gsim a$
\footnote{The choice $n_0P(k_d) \sim 1$ means that, in real space, the
  normalised ``theoretical'' mass variance $\sigma^2(R)$ in spheres of radius
  $R$ (i.e. that corresponding to the model with PS $P(k)$) is of order unity
  at the inter-particle distance. This follows from the fact that, for these
  model PS, one has $\sigma^2(R) \sim k^3 P(k)$, with $k \sim R^{-1}$. Thus
  $\sigma^2(a) \sim k_d^3 P(k_d) \sim n_0P(k_d)$.}.

In our explicit examples of the construction of $S(k)$ we will make
the simple choice $D(k)= 1-e^{-k^2/2k_d^2}$, which evidently has the
required asymptotic properties. It is important to note that we have
not shown that the $S(k)$ then given by Eq.(\ref{prescription}) and such a
choice of $D(k)$ is necessarily the PS of a real discrete distribution
\footnote{For a continuous SSP with finite variance it suffices that
the PS be a positive function with the appropriate convergence
properties at small and large $k$ (to make its integral finite). For
the discrete case the existence conditions on $S(k)$ are, apparently,
not known. Note, in particular, that it is not clear whether
there are intrinsic limits on the small $k$ behaviour of $S(k)$.  In
the case that such limits are established an elegant choice for $D(k)$
would be one giving this limiting small $k$ behaviour.  One would then
have that the ``discretisation'' of a uniform continuous distribution
(i.e.  $P(k)=0$) would be the (or one of the class of) most uniform
possible discrete distributions.}. Indeed it is easy to see that the
ansatz for $S(k)$ may be unrealizable in a discrete distribution: we
have noted that the two-point correlation function $h(r)$ of the
discrete distribution must satisfy by definition $h(r)\geq -1 $.
Taking Eq.(\ref{corrfn-condition2}), it is not difficult to verify that
this condition places an upper bound on $k_d$, of order the inverse 
of the average inter-particle distance \footnote{The exact
numerical value for the bound in the case $D(k)= 1-e^{-k^2/2k_d^2}$
will be given at the appropriate point below.}. Physically it is very
reasonable that such a bound arises: taking $k_d$ larger than the
inverse of the inter-particle separation one is requiring the discrete
distribution to mimic the correlation properties of the continuous
model in a regime where the intrinsic difference in the nature of the
distributions is important.

\section{SEMI-ANALYTIC DETERMINATION OF THE POTENTIAL: THE HNC}
\label{hnc}

In this section we discuss first the OCP and a standard method
use to determine its correlation properties. The inversion 
of this method to determine the potential which would be
expected to give reasonable desired input correlation properties
is then explained.

\subsection{The standard OCP}

The OCP (for a review, see \cite{baus_1980}) is a system of positive charged
point particles (``ions'') interacting through a Coulomb (i.e. repulsive
$1/r$) potential, and embedded in a uniform (rigid, non-dynamical) negatively
charged background \footnote{Changing the sign of the gravitational
  interaction in a cosmological N-body simulation of gravity gives in fact
  just this model.  As mentioned above, such a system is sometimes used to
  generate the ``pre-initial'' configuration (instead of the lattice) for the
  standard algorithm of generating IC.  From the results quoted below one can
  infer the statistical properties of these configurations. Note that what we
  are considering here is quite different: the modification of the interaction
  in the OCP in order to produce {\it directly} the desired IC.}. The latter
gives overall charge neutrality, and a high degree of stability to the system.
The system exhibits two phases at thermal equilibrium, a fluid phase and a
solid phase.  We will treat it always at densities and temperatures where it
is in the fluid phase. In this range of densities and temperature it can be
considered as completely classical.

The equilibrium thermodynamics of the OCP is determined by a single parameter,
and not by its temperature and density independently. Because of the
scale-free nature of the power-law interaction potential, there are only two
characteristic length scales. One is specified by the number density, and is
conventionally taken to be the ``ion-sphere'' (or simply ionic) radius $a$
defined by \be
\label{ion-sphere-radius}
a=\left(\frac{3}{4\pi n}\right)^{1/3}
\ee
where $n=N/V$ is the number density of the $N$ points in
a volume $V$. The other scale is given by the distance at
which the potential is of order the mean thermal kinetic
energy. It is the dimensionless ratio of these two scales
which parametrises the one dimensional phase space 
of the system at thermal equilibrium.
Conventionally this parameter is taken to be
\be
\label{gamma}
\Gamma=\beta(Ze)^2/a. 
\ee
where $\beta=1/(k_B T)$ and $Ze$ is the ionic charge.
It is refered to as the ``plasma parameter''
(or simply ``coupling constant''). 

\begin{figure}
\resizebox{8cm}{!}{\includegraphics*{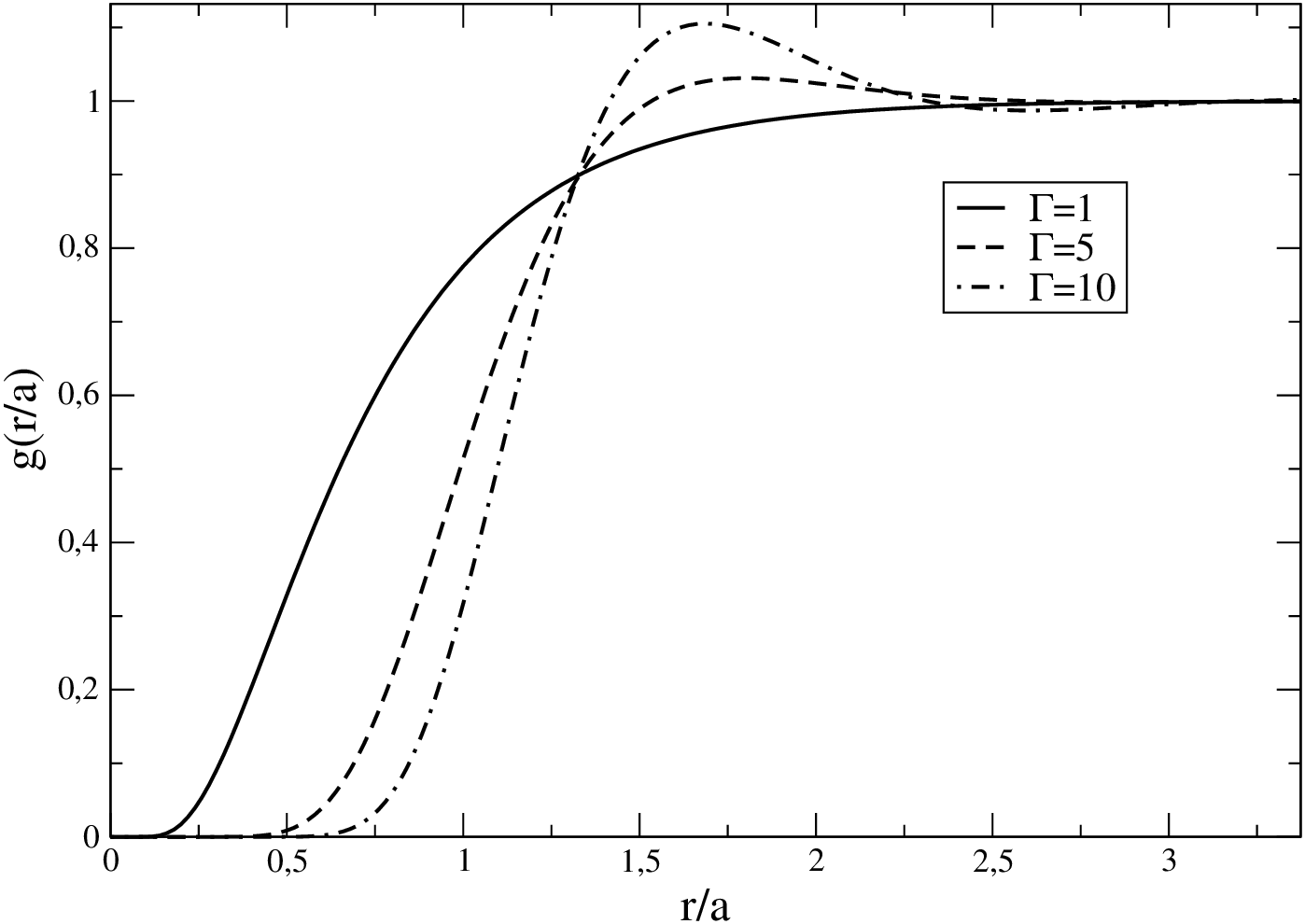}}
\caption{Correlation function of the OCP with interacting Coulomb potential for different temperatures (recall that $\Gamma\sim 1/T$).\label{ocp_real}}
\end{figure}

\begin{figure}
\resizebox{8cm}{!}{\includegraphics*{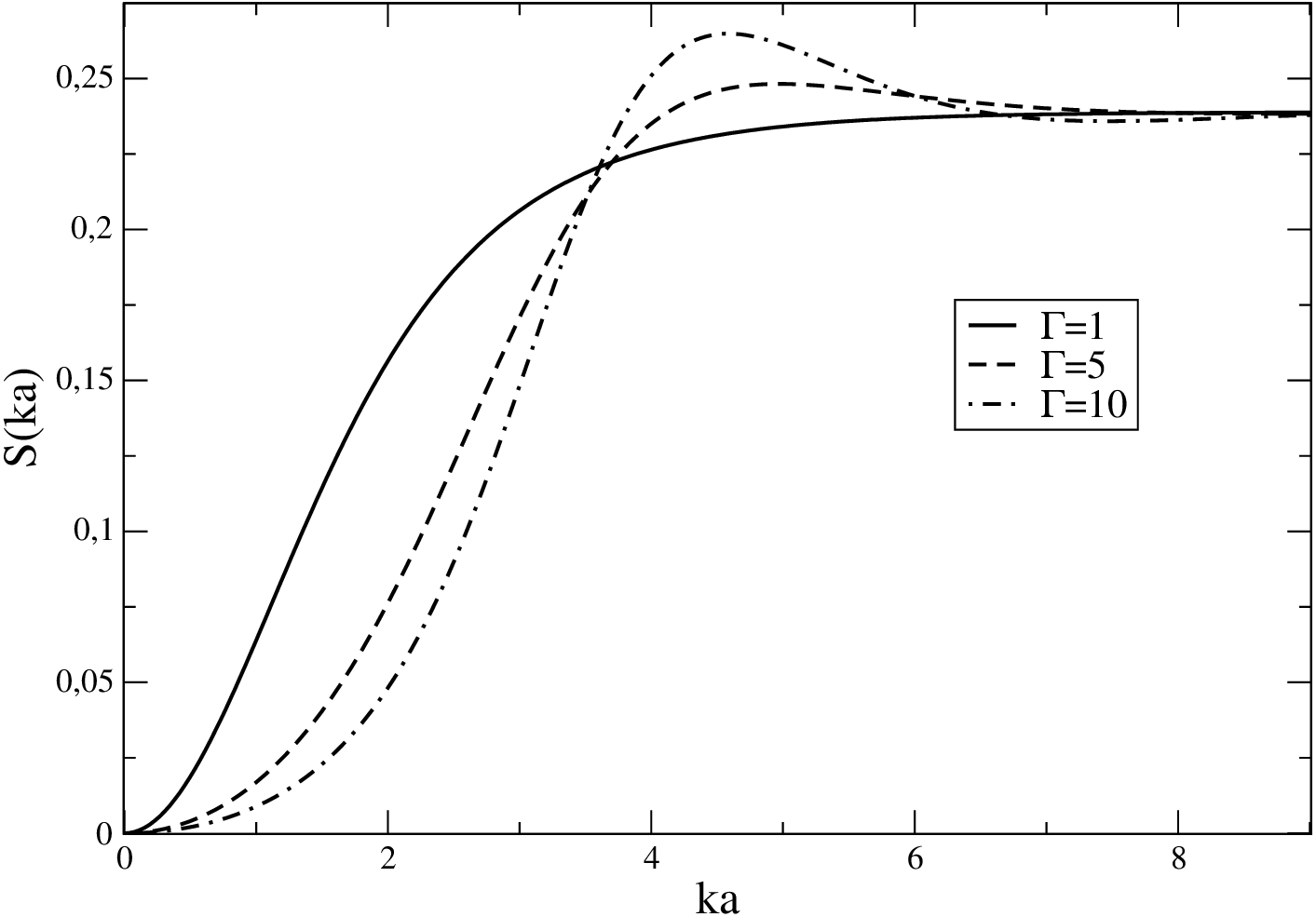}}
\caption{Power spectrum of the OCP with interacting Coulomb potential.
Note that this is a linear-linear plot. \label{ocp_fourier}}
\end{figure}

Numerical calculations are necessary to compute reliably
the correlation properties of the OCP.
The correlation function $g(r)$ (defined as $g(r)=h(r)+1$)
and PS $S(k)$ are shown
in Figs. \ref{ocp_real} and \ref{ocp_fourier} for different
values of the coupling $\Gamma$. 

In the weak coupling limit ($\Gamma \ll 1$ i.e. high temperature/low density)
the OCP can be shown to have two point correlations given approximately as
\be
h(r)\equiv \left(g(r)-1\right) 
\sim -\beta\frac{e^{-\kappa r}}{r},\quad \kappa^2=4\pi Z^2 e^2 \beta n.
\ee 
i.e. exponentially decaying anti-correlations. This is simply
a manifestation of screening, and $\kappa^{-1}$ is the 
Debye length: 
\be
\lambda_D=(4\pi Z^2 e^2 n \beta)^{-1/2}=\frac{a}{\sqrt{3\Gamma}}.
\ee
Particles are repelled locally around a given particle in such
a way that the effective potential due to a particle decays
exponentially in this way. At larger values of $\Gamma$ (i.e.
lower temperature/higher density) one sees that the correlation
function develops a ``bump'' at small scales, indicating that
the first neighbour is becoming increasingly localised.
As $\Gamma$ increases further several  ``bumps'' develop
(corresponding to first, second, third neighbours) which
give to the correlation function and PS an oscillatory
structure, fore-shadowing the transition to the ordered
solid phase at $\Gamma\approx 180$ (for more details, see \cite{dewitt}).

By taking the Fourier transform
it is easy to verify that, at small $k$ the PS has the
form
\be
\label{OCP-asymptoticPS}
S(k)\sim\frac{k^2}{4\pi n^2 \beta},\,\mathrm{for}\,k\ll\frac{2\pi}{\lambda_D}.
\ee
The behaviour $S(k\rightarrow 0) \rightarrow 0$ we see 
in all cases is a manifestation
in $k$ space of the very strong suppression of the 
fluctuations at larger scales in the system due to screening. 
The fluctuations in particle number (i.e. in charge) in a 
volume are proportional 
to the {\it surface} of the volume, which is in fact the limiting
behaviour possible for any point distribution \cite{beck}.
This is actually \cite{glass, gsljp-springer} a  property 
shared by the PS of the canonical model in cosmology 
for ``primordial'' fluctuations (the so-called 
``Harrison-Zeldovich'' (HZ) spectrum),  which has the 
different small $k$ behaviour $P(k) \sim k$.

In \cite{gjjlpsl02} it has been pointed out that, for
a modification of the OCP with a repulsive potential
$v(r)$, screening generically leads to the behaviour for small $k$
\be
\label{asympotic_P}
S(k)\sim \mathrm{FT}[v(r)],
\ee 
(where the FT is taken in the sense of a distribution). 
While for the Coulomb potential one obtains 
the behaviour of Eq.(\ref{OCP-asymptoticPS}),
it is evident that by taking instead an interaction
potential $v(r)\sim 1/r^2$, one should obtain 
\be
S(k)\sim\frac{k}{2\pi^2 n^2 \beta},\,\mathrm{for}\,k\ll\frac{2\pi}{\lambda_D}.
\ee

As we will discuss in further detail at the appropriate point
below, cosmological PS which are used as the input for
numerical simulations of structure formation are more
complicated than the simple HZ form. In principle, however,
it seems possible that an appropriate further modification 
of the interaction potential could allow a system at
thermal equilibrium to produce more complicated PS.

\subsection{Determination of $h(r)$ using the HNC equation}

The {\em Hypernetted Chain Equation}, which 
can be derived in a perturbative treatment of the full 
equilibrium thermodynamics (see e.g.\cite{hansen_76}, chapter 5),  
gives a simple relation between the full two point 
correlation function and the interaction potential: 
\begin{equation}
\label{hnc_eq}
h(r)=\exp[-\beta v(r) +h(r)-c(r)]-1 \,.
\end{equation}
The function $c(r)$ is the {\em direct correlation function}.  It
gives the contribution to the correlation coming only from the
interaction of two particles at the given separation (i.e. neglecting
all indirect correlation due to interaction with other particles). It
is related to the full correlation function through the
\textit{Ornstein-Zernike} (OZ) relation
\begin{equation}
\label{ornstein3}
h(r)=c(r)+n \int d 
\vec{r}\,'c(\lvert\vec{r}-\vec{r}\,'\rvert)h(\vec{r}\,').
\end{equation}
Given the potential $v(r)$, Eqs.(\ref{hnc_eq}) and (\ref{ornstein3})
give a closed set of equations for the correlation function $h(r)$,
which can be solved by iteration as follows. Transformed to Fourier
space Eq.(\ref{ornstein3}) is a simple convolution product
\begin{equation}
\label{ornstein_k}
\tilde h(k)=\tilde c(k)+n \tilde h(k) \tilde c(k).
\end{equation} 
which allows one to express $\tilde h(k)$ 
in terms of $\tilde c(k)$ as 
\begin{equation}
\label{algo1}
\tilde h(k)=\frac{\tilde c(k)}{1-n \tilde c(k)}\,.
\end{equation}
It is convenient to define $\gamma(r)=h(r)-c(r)$, 
of which the FT $\tilde\gamma(k)$ is given in terms 
of $\tilde c(k)$ as
\begin{equation}
\label{algo2}
\tilde \gamma(k)=\frac{\tilde c(k)}{1-n \tilde c(k)}-\tilde c(k).
\end{equation}

We start with a first guess for $c(r)$, denoted $c_0(r)$.  One can
take $c_0(r)=0 \,\,\,\forall r$, or, for example, $c_0(r)\simeq-\beta
v(r)$, which is the solution to the HNC equation Eq.(\ref{hnc_eq}) at
leading order in expansion of the exponential.  We can then use a Fast
Fourier transform (FFT) to calculate $\tilde c_0(k)$, which then gives
$\tilde \gamma_0(k)$ through (\ref{algo2}). With an inverse FFT we
find $\gamma_0(r)$, and then use the HNC equation Eq.(\ref{hnc_eq}) to
compute $c_1(r)$ (using $\gamma_0(r)$ in the exponent to obtain
$c_1(r)+\gamma_0(r)$ on the left hand side).  The iteration process
then proceeds with the computation of $\gamma_i(k)$ with
(\ref{algo2}). To ensure convergence, successive approximations on
$\gamma(r)$ need to be taken, so the $i$th input is mixed linearly
with the precedent one:
\begin{equation}
\gamma_i'(r)=\alpha \gamma_{i-1}(r)+(1-\alpha)\gamma_i(r)
\end{equation}
where $0<\alpha<1$. The new $\gamma_i'(r)$ is substituted 
in equation (\ref{algo2}) to get $c_{i+1}(r)$ and
so on.  In the simulations described below we 
took $\alpha=0.5$ which gives rapid convergence
(less than one hundred iterations were necessary
in all cases). If there are problems with convergence
(which can occur e.g. at larger densities) a value 
of $\alpha$ closer to $1$ is taken.

There is one additional elaboration of this method which
is necessary when the potential is long-range, as it is
for the case of the standard OCP and all the modifications
which we will consider in what follows. Since 
\begin{equation}
\label{c_k}
\tilde c(k)=\frac{\tilde h(k)}{1+n \tilde h(k)}=
\frac{1}{n}\left(1-\frac{1}{n_0 S(k)}\right)
\end{equation} 
we have that $\tilde c(k)$ diverges for $k \to 0$, which is
problematic numerically. This is dealt with in an analogous manner to
that described in the Sect. \ref{mol_dyn} below for the calculation of
the force by the Ewald sum. One breaks $c(r)$ into the sum of a
short-range part $c_s(r)$, with an analytic FT at $k=0$, and a long
part $f(r)$, which contains the divergence in the FT. A typical long
range part is $v(r)\mathrm{erf}(\eta r)$ or $v(r)(1-\exp(-\eta r))$, where
$\eta$ is a free positive parameter (on which the final result does not
depend). The total correlation function $h(r)$ has no divergence, and
thus $\gamma(r)$ is divided in the same way,
$\gamma(r)=\gamma_s(r)+f(r)$, with $\gamma_s(r)=h(r)-c_s(r)$. The
potential likewise is separated into a short and long range part
$\beta v_s(r)=v(r)+f(r)$, so that the HNC reads
\begin{equation}
h(r)=\exp[-\beta v_s(r)+\gamma_s(r)]-1.
\end{equation}
When we compute Eq.(\ref{algo2}) we use the FT of the long-range part
of the potential: 
\be 
\label{algo2_regul}
\tilde \gamma_s(k)=\frac{\tilde c_s(k)+\tilde
f(k)}{1-n (\tilde c_s(k)+\tilde f(k))}-\tilde c_s(k).  
\ee 
All the computations are then done as described above but with $c_s(r)$ and
$\gamma_s(r)$ instead of $c(r)$ and $\gamma(r)$, and using Eq. (\ref{algo2_regul}) 
instead of Eq.(\ref{algo2}).

\subsection{Inversion of the HNC}

It is simple to use the HNC equation 
in the inverse direction i.e. to determine an interaction
potential $v(r)$ which should give at thermal equilibrium
desired two-point correlation properties, for we have 
directly from Eq.(\ref{hnc_eq}) that
\begin{equation}
\label{inv_hnc}
\beta v(r)=h(r)-c(r)-\ln[h(r)+1].
\end{equation}
Starting from an input model specified by a given PS $S(k)$
we need just to calculate $h(r)$ and $c(r)$ (using 
the OZ relation Eq.(\ref{ornstein_k})). This can most
conveniently be done using FFTs. 

As noted above, when we treat the case of a PS with
$S(k \rightarrow 0)=0$, characteristic of a 
long-range interaction potential, we have a divergence at $k=0$
in $\tilde c(k)$. Just as in the direct use of the HNC 
we deal with this numerically by dividing $\tilde c(k)$
into two parts. The short-range part, which is regular
at $k=0$, can be taken to be 
\begin{equation}
\label{c_reg}
\tilde
c_s(k)=\frac{1}{n_0}\left(1-\frac{1}{n_0 S(k)}+\frac{\mathrm{erfc}(k \eta)}{n_0 S_0(k)}\right).
\end{equation}
where $S_0(k)$ is the functional form of $S(k)$ at small $k$, and as
above, $\eta$ is a parameter on which the final result does not
depend. The subtracted divergent piece is chosen (if possible) so that
it can be Fourier transformed analytically, and the full potential can
thus be reconstructed easily from a determination of the short-range
part of the potential from Eq.(\ref{inv_hnc}) using $c_s(r)$: 
\be
\beta v(r)=\beta v_s(r)-\mathrm{FT}[\tilde{c}_l(k)], 
\ee 
where
$\tilde{c}_l(k)$ is the long-range part of $\tilde{c}(k)$,
which  corresponds to $f(r)$ in the precedent subsection. 

\section{DETERMINATION OF THERMAL EQUILIBRIUM PROPERTIES USING 
MOLECULAR DYNAMICS}
\label{mol_dyn}

The two numerical methods used widely in statistical physics to study
systems at thermal equilibrium are molecular dynamics and 
Monte Carlo simulations. We will discuss the former method, 
in which one evolves numerically the $3N$ classical coupled 
equations of motions of a system of $N$ interacting particles 
in a volume $V$. Finite-size effect are treated using 
periodic-type boundary conditions. We discuss the 
modification of a standard OCP code, developed by one
of us (DL),  to study the reliability of the
(approximate) HNC method described in the previous section.
We are interested therefore in simulating in this way 
a system of  particles interacting through two-body potentials 
expected to generate systems with two-point correlation properties 
like those of  cosmological models. Further, 
if this kind of method is to be used to generate initial conditions 
for N body simulations of gravity, these simulations provide
an instrument for producing the required configurations.

\subsection{Discretisation of the Newton equations}
To discretise the equations of motion we use the Verlet algorithm.
Performing a Taylor expansion of the position of a particle at
times $t+\Delta t$ and $t-\Delta t$ about its position at time $t$,
the position of the \textit{i-th} particle is given to order
$\mathcal{O}((\Delta t)^4)$ by:
\begin{equation}
\vec{r}_i(t+\Delta t)= 2 \vec r(t)-\vec r(t-\Delta
t)+\frac{(\Delta t)^2}{m}\sum_{i=1}^N\vec F_{ij}(t) \,.
\end{equation}
This algorithm, which is historically one of the earliest ones, has
three important properties: it conserves energy very well, it is
reversible (as the Newton equations), and it is simplectic (i.e. it
conserves the phase space volume). More refined algorithms have been proposed
and used, but they often have less good conservation of energy at
large times.  Furthermore, the rapidity of the execution of the
program is not determined by the computation of the new positions but
by the calculation of the forces.

\subsection{Force calculation using the Ewald sum}

In our simulations $N$ particles are placed in a cubic box of size $L$. To
compute the interaction between the particles we apply the {\em image} method
to minimize border effects:
an infinite number of copies of the system is created and the potential is
computed considering not only the particles situated in the original box but
also the particles of all the copies. Then if the particle $i$ has coordinate
$\vec r_i$, its copies will have cordinates $\vec r_i+\vec n L$, where $\vec
n$ is a vector with integer components.   For a power-law interaction potential
$v(r)=r^{-\alpha}$ the potential is then 
\begin{equation}
\label{potl_j}
\phi(\vec r_i)=\sum_{j,\vec n}^{*}\frac{q_j}{\lvert \vec
r_{ij}+\vec n L \rvert ^{\alpha}},
\end{equation}
where $q_j$ is the charge of the particles and the asterisk denotes
that the sum $\vec n=0$ does not include the term $i=j$.  In a
numerical calculation the infinite sum Eq.(\ref{potl_j}) must be
truncated. For $\alpha>3$ the potential is short-range and the
approximation to compute the interaction potential between the $i$ and
$j$ particles by taking only the interaction between $i$ and the
closest image of $j$ is very good. When the potential is long-range
($\alpha<3$) this approximation is no longer good, and indeed the
sum appears to be formally divergent. For the case of the Coulomb
potential, the presence of the neutralising uniform background ensures
that the potential of the infinite periodic system is well defined. A
natural way of writing the sum in an explicitly convergent way taking
this regularisation into account is to separate the potential into a
short range and long range part by introducing a parameter-dependent
damping function $f(\vec r;\alpha)$:
\begin{equation}
\label{damping}
\phi(\vec r_i)=\sum_{j,\vec n}^{*}q_j\left(\frac{f(\vec r_{ij}+\vec nL;\alpha)}{\lvert \vec
r_{ij}+\vec n L \rvert ^{\alpha}}+\frac{1-f(\vec
r_{ij}+\vec nL;\alpha)}{\lvert \vec r_{ij}+\vec n L \rvert
^{\alpha}}\right)\,.
\end{equation}
The first term on the r.h.s of Eq.(\ref{damping}) is short-range
and the second term is long-range. The procedure used in the
Ewald summation method is to compute the first term
in real space and the second in Fourier space. If the parameter
$\alpha$ is appropriately chosen the real part converges well
taking only the sum over the closest image, and the  
part of the sum in Fourier part is rapidly convergent. 
Physically the first term corresponds to a smearing of
the original distribution, and the second term to the original point
distribution sorrounded by a countercharge smeared distribution. Of
course the sum of the two terms yields the original particle
distribution. We write the potential energy then as:
\begin{equation}
\label{energy}
\phi=\phi_{\vec r}^{(s)}+\phi_{\vec k}^{(l)}.
\end{equation}
Further it is convenient to separate out the zero mode
in the long range part, writing
\begin{equation}
\phi_{\vec k}^{(l)}=\phi_{\vec k=0}^{(l)}+\phi_{\vec k\neq 0}^{(l)}.
\end{equation}
The function $f(\vec r;\alpha)$ is chosen in the
Ewald summation so that $\phi_{\vec r}^{(s)}$ and 
$\phi_{\vec k\neq 0}^{(l)}$ are both rapidly
convergent, and with a known analytical expression
for its Fourier transform. The value of the term 
$\vec k=0$ depends on how precisely the  
infinite sum in Eq.(\ref {potl_j}) is defined,
and, as we will see further in particular examples, it is equal to zero
in the presence of the  background because of the charge neutrality.
This method of evaluating the potential energy using 
the Ewald Sum has been generalised for generic $r^{-\alpha}$ potentials
\cite{wu01}, and for a Yukawa potential \cite{caillol02}. In principle
it may be used for other potentials. Note in particular that the 
Ewald method  is applied to sum  the long-range part of the 
potential: it remains valid if one introduces any additional
short-range potential which can be absorbed in
$\phi_{\vec r}^{(s)}$ without modification of 
$\phi_{\vec k}^{(l)}$. We now give more detail first on 
its implementation for the standard OCP, and then for the
potentials which will interest us here. 

\subsubsection{Case 1: Coulomb interaction}

The $f(\vec r;\alpha)$ function is usually chosen to be
\begin{equation}
f(\vec r;\alpha)=\mathrm{erfc}(\alpha\lvert\vec r+\vec nL\rvert)
\end{equation}
where erfc is the complementary error function, $\mathrm{erfc}(x)\equiv
1-2/\sqrt{\pi}\int_0^x dt\exp(-t^2)$. It is equivalent to smearing
the charge distribution to obtain 
\begin{equation}
\label{smear}
\rho(\vec r)=\sum_{j=1}^N\sum_{\vec n} q_j
\exp\left(-\alpha\lvert\vec r-(\vec r_j +\vec n
L)\rvert^2\right)
\end{equation}
and introducing in Fourier space the original distribution plus the opposite smeared distribution.
With this choice the short-range interaction energy is given by
\begin{equation}
\label{ewald1}
\phi_{\vec r}^{(s)}(\vec r_i)= \sum_{j=1}^N \sum_{\vec
n}q_j \frac{\mathrm{erfc}(\alpha\lvert\vec r_{ij}+\vec n L
\rvert)}{\lvert\vec r_{ij}+\vec n L\rvert} \,,
\end{equation}
and the long-range part by
\begin{equation}
\label{ewald2}
\phi_{\vec k\neq 0}^{(l)}(\vec r_i)=\frac{4\pi}{L^3} \sum_{j=1}^N
\sum_{\vec k \neq 0}q_j \frac{1}{k^2}
\exp\left(\frac{-k^2}{4\alpha^2}\right)\cos(\vec k \vec r_{ij}).
\end{equation}
The $\vec k=0$ term is thus well defined only when the
constraint  
\begin{equation}
\frac{1}{L^3} \sum_{j=1}^N q_j=0 
\end{equation}
is satisfied i.e. only for a neutral distribution of 
charge. In the present case this is imposed by the 
uniform rigid negatively charged background assumed.

An appropriate choice of $\alpha$ is 
$\alpha\approx 5.6/L$, where $L$ is the size of the box.
This gives good
convergence in both (\ref{ewald1}) and (\ref{ewald2}), 
i.e. it includes only the first term
$\vec n=0$ in the first equation and not too many $\vec k$ in
the second.

\subsubsection{Case 2: $1/r^2$ potential}
With a potential $1/r^2$ a convenient choice 
for $f(\vec r;\alpha)$ is \cite{wu01}:
\begin{equation}
f(\vec r;\alpha)=\exp(\alpha^2\lvert\vec r+\vec nL\rvert^2).
\end{equation}
The short-range part of the energy is
\begin{equation}
\phi_{\vec r}^{(s)}(\vec r_i)= \sum_{j=1}^N \sum_{\vec
n} q_j \frac{\exp(\alpha^2\lvert\vec r_{ij}+\vec n L
\rvert^2)}{\lvert\vec r_{ij}+\vec n L\rvert^2}
\end{equation}
and the long-range part
\begin{equation}
\phi_{\vec k\neq 0}^{(l)}(\vec r_i)=\frac{2\pi^2}{L^3}\sum_{j=1}^N
\sum_{\vec k \neq 0}q_j \frac{1}{k}
\mathrm{erfc}\left(\frac{k}{2\alpha}\right)\cos(\vec k \vec
r_{ij}).
\end{equation}
We will use the same value for $\alpha$ as in the Coulomb
case. With this value of $\alpha$ the real part still converges
rapidly and the Fourier part is much more rapidly convergent.

\subsubsection{Case3: inversion of HNC}
In what follows we will wish to simulate the molecular 
dynamics of particles interacting through the potential
determined by the inversion of HNC equation as described
in the previous section. As discussed in 
Sect. \ref{discretisation}, the small $k$ behaviour 
of cosmological PS (the HZ spectrum of perturbations), 
requires a long-range $1/r^2$ potential. In the
determination of the full potential through the inversion
of the HNC, this piece is separated out by construction
and the result is written as a sum of it and the short-range 
part subsequently determined. Taking the long-range
part that comes from the subtracted divergence on
the r.h.s. of Eq.(\ref{c_reg}), the long-range
part is thus in this case
\begin{equation}
\phi_{\vec k\neq 0}^{(l)}(\vec r_i)= \frac{1}{n_0 L^3\beta}\sum_{j=1}^N
\sum_{\vec k \neq 0} q_j \frac{1}{n_0 S_0(k)}
\mathrm{erfc}\left(\frac{k}{2\alpha}\right)\cos(\vec k \vec
r_{ij})\,,
\end{equation}
where $S_0(k)=Nk$ gives the small $k$ behaviour of $S(k)$.
The real part of the potential is then:
\begin{equation}
\phi_{\vec r}^{(s)}(\vec r_i)=\frac{\exp(-\alpha^2 r^2)}{2\pi^2 {n_0}^2 N \beta r^2}.
\end{equation}
Note that the parameter $\alpha$ in the Ewald sum needs to
have the same numerical value as the parameter $\eta$ in the HNC.

\section{GENERATION OF DISCRETISATIONS OF COSMOLOGICAL SPECTRA}
N-body simulations of the formation of structure in the distribution
of matter at large scales start from an initial time which is 
``recent'' in terms of cosmological history. The universe has 
entered the phase in which its energy density is dominated
by massive particles, and the evolution of perturbations in
the distribution of these particles at the scales considered
is well approximated by Newtonian gravity. The fluctuations 
at this initial time
are still of small amplitude at the relevant physical 
scales, and the simulation follows this evolution through to
today when very high amplitude fluctuations have formed
at scales comparable to those on which they are observed to
exist today. These initial conditions for simulations are
generically Gaussian in current cosmological models, and
thus fully specified by their PS. This PS is the result
of the evolution up to this time, which can be calculated
precisely in a given model (and depends on the various
parameters characterising it) of the ``primordial'' 
fluctuations, which usually have the form given by
so-called ``scale-invariant'' fluctuations\footnote{The term
is here used in a sense different to that in the context
of statistical physics, where is applies to a wide class
of fluctuations (and PS) manifesting properties of invariance under
transformations of scale. See \cite{gsljp-springer}. }. 
Because the fluctuations evolve in a non-trivial way 
for a finite time (until the time of ``equality'',
after which matter dominates over radiation) the 
resultant PS corresonds to the ``primordial'' spectrum
$P(k) \sim k$ only up to a characteristic wave-number $k_t$, 
above which it ``turns over'' to a different behaviour, with
a PS which decreases as a function of $k$ but with
a functional behaviour which depends on the model.
We will consider here the class of ``cold dark matter''
(CDM) models which are those currently favoured as viable
models to explain the diverse observations of large scale
structure. In N body simulations the initial
spectrum is usually taken in this case as given by a
simple functional fit to the results of a numerical 
determination of the PS (see e.g. \cite{jenkins98}):
\begin{equation}
\label{P_CDM}
P(k)=
\frac{{\cal N}(z) k}{\left(1+(aq+(b q)^{3/2}+(c q)^2)^{\nu}\right)^{2/\nu}}
\end{equation}
where $q=k/\Lambda$ is a rescaling of $k$ by a dimensionless parameter
$\Lambda$ which depends on the parameters of the CDM model
($\Lambda=0.21$ for ``standard'' CDM), and $\nu=1.13$. 
In units of $h^{-1}$ Mpc, where
$h$ is the Hubble constant today in units of $100$ km/s/Mpc, one has
$a=6.4$, $b=3$ and $c=1.7$. The factor ${\cal N}(z)$ gives the overall
normalisation of the PS, which is a function of the initial red-shift
$z$ (for a red-shift chosen in the matter dominated era, during which
the fluctuations are, to a very good approximation,  simply amplified 
in the same way at all scales.) It is in principle fixed by the 
amplitude of fluctuations
measured in the cosmic micro-wave background (CMB), and is often
specified by giving the value of $\sigma_8$, which is the normalised 
mass variance in a sphere of radius $8 h^{-1}$ Mpc calculated from 
the PS when the model is extrapolated linearly to today  (i.e. $z=0$).

The PS thus shows the HZ form at small $k$, reaches a
maximum at $k_t\approx 0.2{\mathrm{(h^{-1} Mpc)^{-1}}}$ and then
interpolates between approximate power-law behaviours
$P(k) \sim k^n$ with $n$ varying from $n \approx -1$ to an 
asymptotic (large $k$) value of $n=-3$
\footnote{To ensure integrability (and the existence of its Fourier
transform) it is strictly necessary to add an ultraviolet cutoff. In
practice this cut-off is usually not made explicit and the Nyquist
frequency acts as the effective cut-off in the discretised model.  See
\cite{jm04} for further detail.}. In practice here we will not work,
for our simulations of molecular dynamics, with the full PS described
in Eq.(\ref{P_CDM}): our simulations are of a size which does not
allow us to resolve the numerous different scales in this
expression.  We use instead a simplified version of this PS which
retains its essential qualitative features:
\begin{equation}
\label{P_simpl}
P(k)=\frac{{\cal N} k}{1+(A k)^\alpha\exp{(k/k_c)}} \,,
\end{equation}
with the maximum $k_t$ chosen well inside the simulation box. 

Following the discussion in Sect. \ref{discretisation}
we seek to produce a discrete distribution with
PS  $S(k)$ given by Eq.~(\ref{prescription})
with  
\begin{equation}
\label{smooth_prescription}
D(k) = \left(1-e^{-k^2/2k_d^2}\right).
\end{equation}

We note that, with this choice for the function $D(k)$, the upper bound
on $k_d$, taking in Eq.(\ref{corrfn-condition2}) $\xi_c(r)=0$ and
using the condition $h(r)\geq-1$, is:
\begin{equation}
\label{value-k_d}
k_d\leq \sqrt{2\pi}(n_0)^{1/3} \approx 1.55/a,
\end{equation}
where we have used the definition of $a$ given 
in Eq.~(\ref{ion-sphere-radius}). By
increasing $n_0$ sufficiently one can represent the 
continuous model up to a desired $k$. 
\begin{figure}
\resizebox{8cm}{!}{\includegraphics*{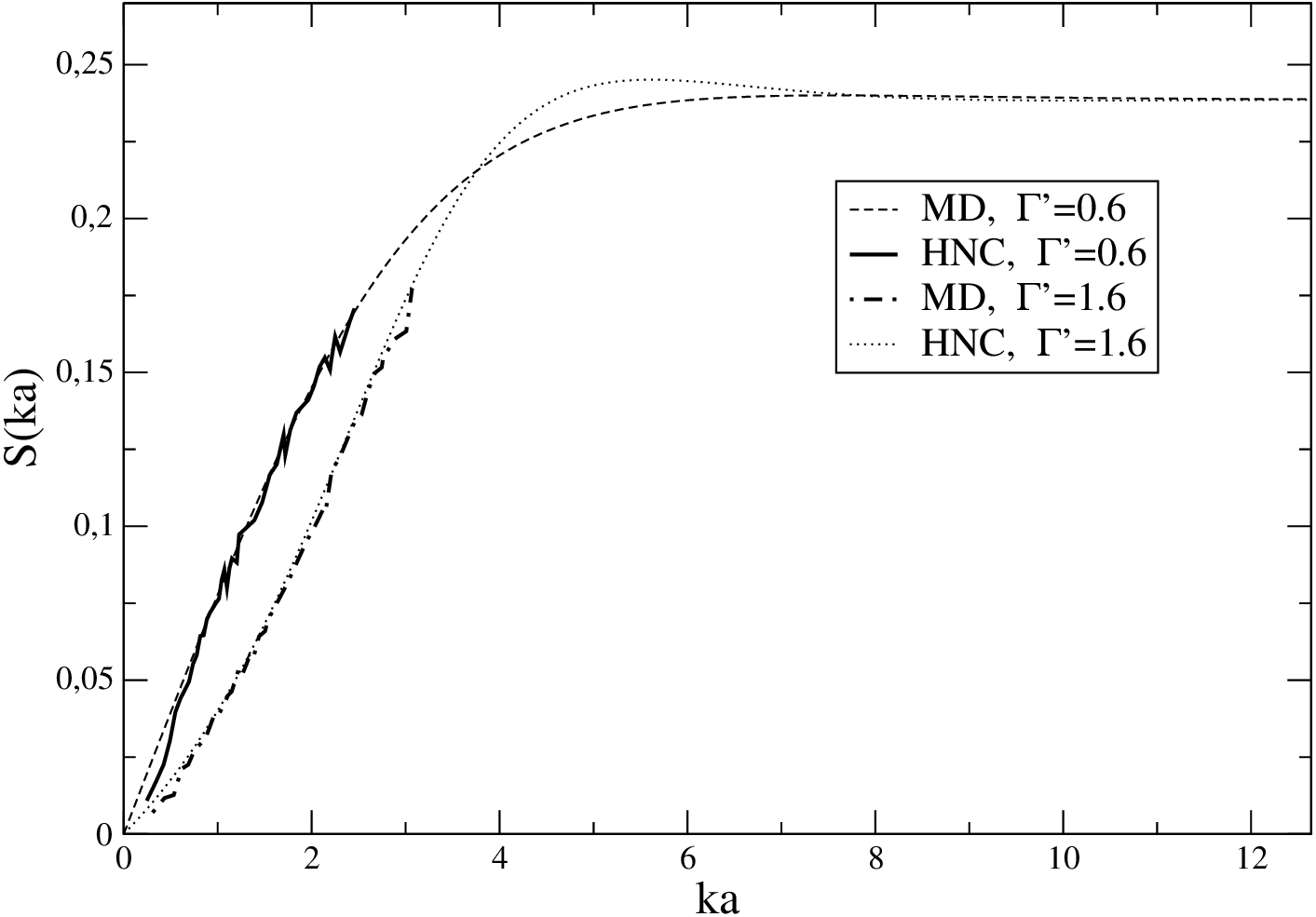}}
\caption{The PS of a $1/r^2$ OCP for two different
values of the coupling parameter $\Gamma'$. Excellent
agreement is observed between the predictions from 
the HNC and MD in the range where they
overlap. For the weak coupling case the HZ form for
the PS $S(k) \propto k$ is clearly evident.
The units are normalised to the ionic radius a.
Note that the plot is on a linear-linear scale.
\label{kkfourier}}
\end{figure}
\begin{figure}
\resizebox{8cm}{!}{\includegraphics*{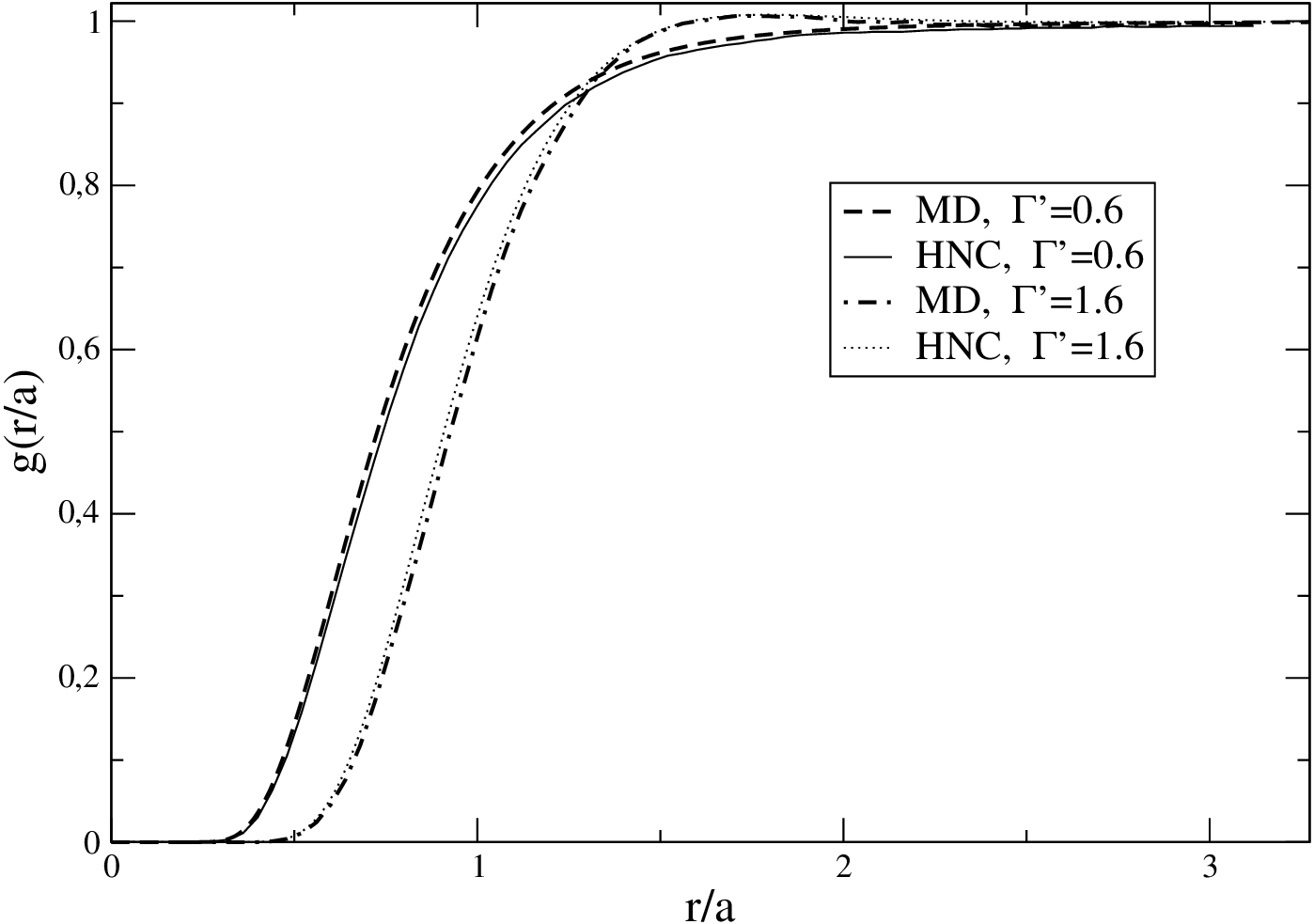}}
\caption{The correlation function in real space
for the $1/r^2$ OCP for the same cases
as in the previous figure. For the smaller coupling
one has anti-correlation at all scales ($g(r) < 1$)
while for the larger coupling one sees, just as in
the standard OCP, the correlation appear with the 
first neighbour (which becomes more localised 
as the temperature is lowered).\label{kkreal}}
\end{figure}

In the first subsection below we will present an example of 
a HZ spectrum generated with a simple $1/r^2$ potential. In
the following subsection we present the method using the 
simplified PS of Eq.(\ref{P_simpl}), while in 
the last subsection we give the potential which should
allow the generation of the ``realistic'' cosmological PS 
of Eq.(\ref{P_CDM}).

\subsection{The HZ spectrum} 

We consider just the ``primordial'' part of the
PS with the HZ behaviour $P(k) \sim k$. As mentioned
in Sect. \ref{hnc}, it has been shown in \cite{gjjlpsl02}
by a simple screening argument that one expects to obtain
such a PS at small $k$  in a modified OCP with interparticle
potential $v(r)=1/r^2$. To verify this expectation we
have used both the HNC and molecular dynamics as described
above. In Fig. \ref{kkfourier} the results for the PS
are given for each case, and in Fig. \ref{kkreal} the 
correlation function in real space. Because the potential
is still a pure power-law the phase space is, as for
the standard OCP, one dimensional and may be 
characterised by a single dimensionless parameter 
analogous to that for the OCP. We make the obvious 
generalisation of the definition in Eq.(\ref{gamma}):
\begin{equation}
\Gamma'=\beta(Ze)^2/a^2.
\label{gamma-prime}
\end{equation}

The results from the HNC are valid in the infinit
volume limit and show very good agreement with the 
prediction of the asymptotic form for both
the PS and the correlation function given in
\cite{gjjlpsl02}. The range of these behaviours
is, as expected, greater for smaller values of 
the coupling, and the linearity of the PS in
particular is clearly visible in this case.
We have checked also that one recovers the
characteristic behaviour of the correlation
function at large scales $(g(r)-1) \sim -1/r^4$,
which is also that of cosmological models
with this PS (see \cite{gsljp-springer,glass}).

The simulations of molecular dynamics were performed in the
micro-canonical ensemble with the methods described in
Sect. \ref{mol_dyn} above, with N=4000 particles \footnote{This is the
number of particles which can be simulated on an ordinary PC for a
reasonable simulation time (a few hours).}.  
This corresponds to a
simulation box with side of length 
$L=\left(\frac{4 \pi}{3} N \right)^{\frac{1}{3}}a
\approx 25.6$. Over this limited range very
good agreement is seen with the results from the HNC in all cases,
with some remaining statistical fluctuations. The units of time used
in the simulations is $\tau=\sqrt{3}\omega_p^{-1}$ with
$\omega_p^2=4\pi n_0 (Ze)^2/m$. To ensure good conservation of energy
we have used a time increment of typically $\Delta t\sim
10^{-2}\tau$, which leads to fluctuations of $\sim 10^{-7}$ in the
energy. The system evolves for $10^5\tau$ times steps, at
which point it has reached thermal equilibrium. Then the PS and
correlation functions are computed over many realizations of the
system. By the ergodic principle this is equivalent to performing an
ensemble average. Each realization is thus a configuration of the system
at each time step. We compute the average in all the simulations over
$50000$ time steps, which leaves only very small 
fluctuations about the average.

\subsection{CDM-type spectra: simple model}

Let us now consider the spectrum (\ref{P_simpl}).
We have seen that the small $k$ part of the spectrum 
can indeed be produced by a repulsive $1/r^2$ potential. 
We will now use the inversion of the HNC to determine the
short range potential which needs to be added to produce 
the modification at larger $k$ described by Eq.~(\ref{P_simpl}).

This input continuous model PS contains four free parameters.
We define $S(k)$, the PS  of the discretisation, using  
Eq.~(\ref{prescription}) with  $D(k)$ as in 
Eq.~(\ref{smooth_prescription}), and
$k_d$ expressed in terms of $a$ by its bounding value 
in Eq.~(\ref{value-k_d}). In discretising we thus introduce 
just {\it one independent length scale}, fixed by the particle 
density $n_0$, and which we identify by convention with
the ionic radius $a$ defined in Eq.~(\ref{ion-sphere-radius}). 
To fix the model being discretised we must specify $S(k)$ 
by giving the three dimensionful quantities (${\cal N}$, $A$ 
and $k_c$) in units of $a$.
The parameter $\alpha$ in 
Eq.~(\ref{P_simpl}) fixes the slope of the power 
spectrum $P(k) \sim k^{1-\alpha}$ 
beyond the ``turn-over'' from the small $k$
behaviour $P(k) \sim k$. As discussed above relevant 
values are in the range $2<\alpha<4$, and we will
consider here the case $\alpha=3$ 
(i.e. $P(k) \sim k^{-2}$ beyond the turn-over).

The cut-off parameter $k_c$ is required
simply to make the theoretical one-point variance 
(given by the integral of the PS) finite.
We will take it, however, to be much
larger than $1/a$, so that it will play essentially
no role in what follows other than to regularise
Fourier transforms at small separations (e.g. 
in determining the theoretical correlation 
function $\xi_c(r)$ associated to Eq.~(\ref{P_simpl})).  
The numerical results  below used
$k_c =2.7/a$.

We thus have essentially just the two parameters   
$A$ and ${\cal N}$ to specify in terms of $a$.
The former fixes the scale of the maximum (or
``turn-over) in the PS: 
\be 
k_t\simeq\frac{1}{(\alpha-1)^{1/\alpha}}\frac{1}{A}\,. 
\ee 
When we simulate the molecular dynamics we have a finite
box of side 
$L=\left(\frac{4 \pi}{3} N \right)^{{1}/{3}} a $,
and thus a finite range, from the inter-particle distance
to the box size, in which we can attempt to represent the 
theoretical correlation properties. In $k$ space 
this is the range $k_f < k < k_d$, where  $k_f=2\pi/L$ 
is the fundamental mode and $k_d$ the $k$-space scale
beyond which $S(k)$ necessarily deviates from the input
$P(k)$. We thus choose $A$ so that $k_t$ falls in this 
range. Once this choice is made ${\cal N}$ then fixes the 
overall normalisation.
As discussed in Sect.~\ref{Determination of the PS of a new discretisation},
the relevant dimensionless parameter to characterise the
amplitude of the fluctuations is $n_0P(k_d)$.
The natural choice, which we will use, is 
$n_0P(k_d)\approx 1$ i.e. the theoretical power
is roughly at the Poisson level at the inverse of the
interparticle distance. 

\begin{figure}
\resizebox{8cm}{!}{\includegraphics*{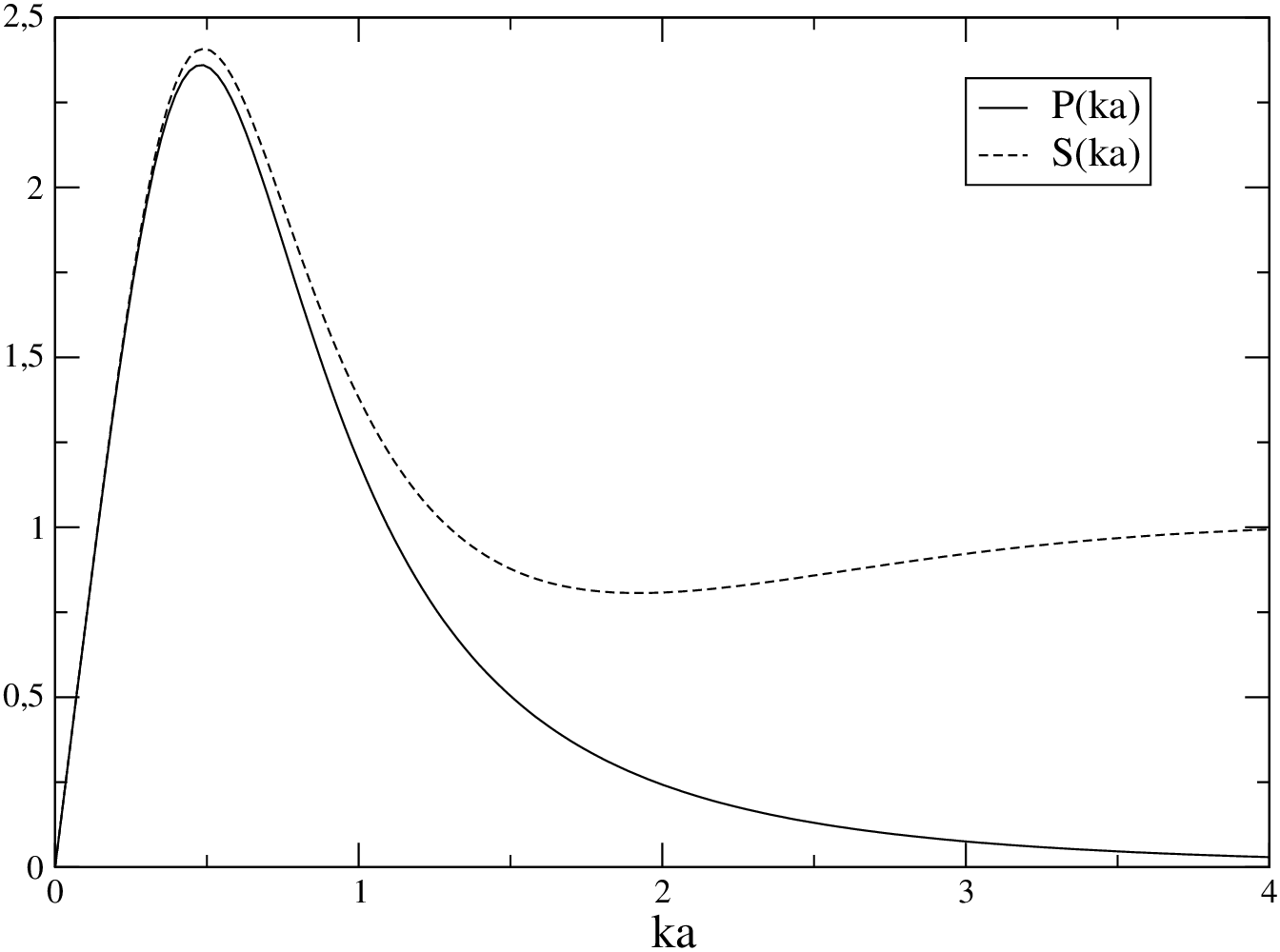}}
\caption{The theoretical PS $P(k)$ and its discretisation
$S(k)$, for the parameters choices given in the text.
They differ as described by the chosen smoothing function $D(k)$,
for $ka \gsim 1$, with $S(k)$ going asymptotically to $1/n_0$, where
$n_0=3/(4 \pi a^3)$ is the particle density for the discretisation.
\label{SandP}}
\end{figure}

\begin{figure}
\resizebox{8cm}{!}{\includegraphics*{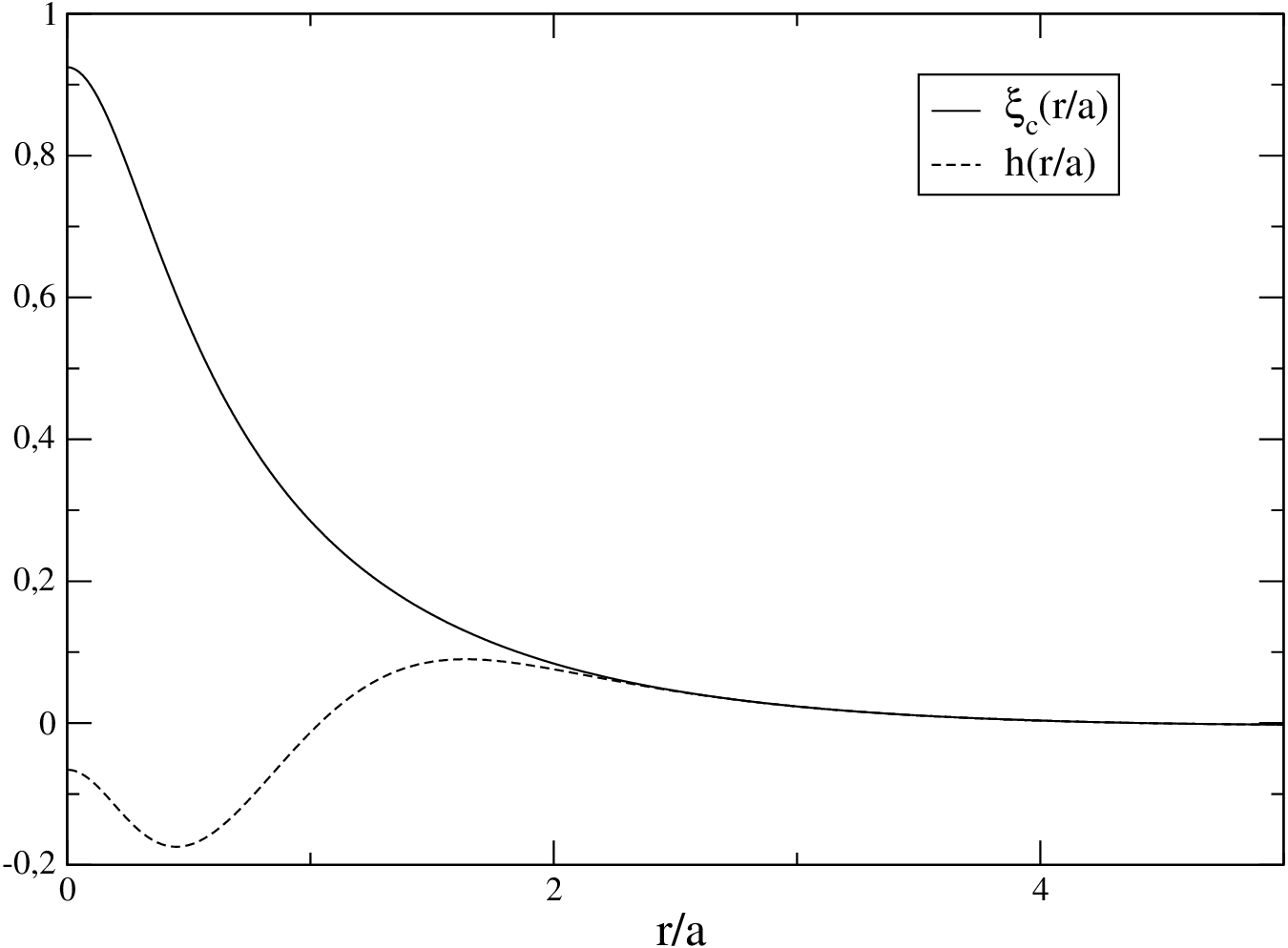}}
\caption{The theoretical input two-point correlation 
function  $\xi_c(r)$ and that of the corresponding  
discretisation $h(r)$. 
\label{gandxi}}
\end{figure}

We will consider here a simulation of $8788$ particles
i.e. $L\approx 33.3a$ and $k_f \approx 0.19/a$. We 
choose $k_t=0.52/a$ so that we have a small range of wavenumbers 
in which the PS is linear in $k$ inside
the box. For the chosen value of $\alpha=3$ this
gives $A\approx 1.51 a$. Finally we 
take ${\cal N}=30a^4$ which corresponds to 
$n_0 P(k_d)=0.47$. In Fig.~\ref{SandP}
the pair $P(k)$ and $S(k)$ corresponding to these values 
are shown, and in Fig.~\ref{gandxi} the corresponding 
correlation function pair $\xi_c(r)$ and $h(r)$.

\begin{figure}
\resizebox{8cm}{!}{\includegraphics*{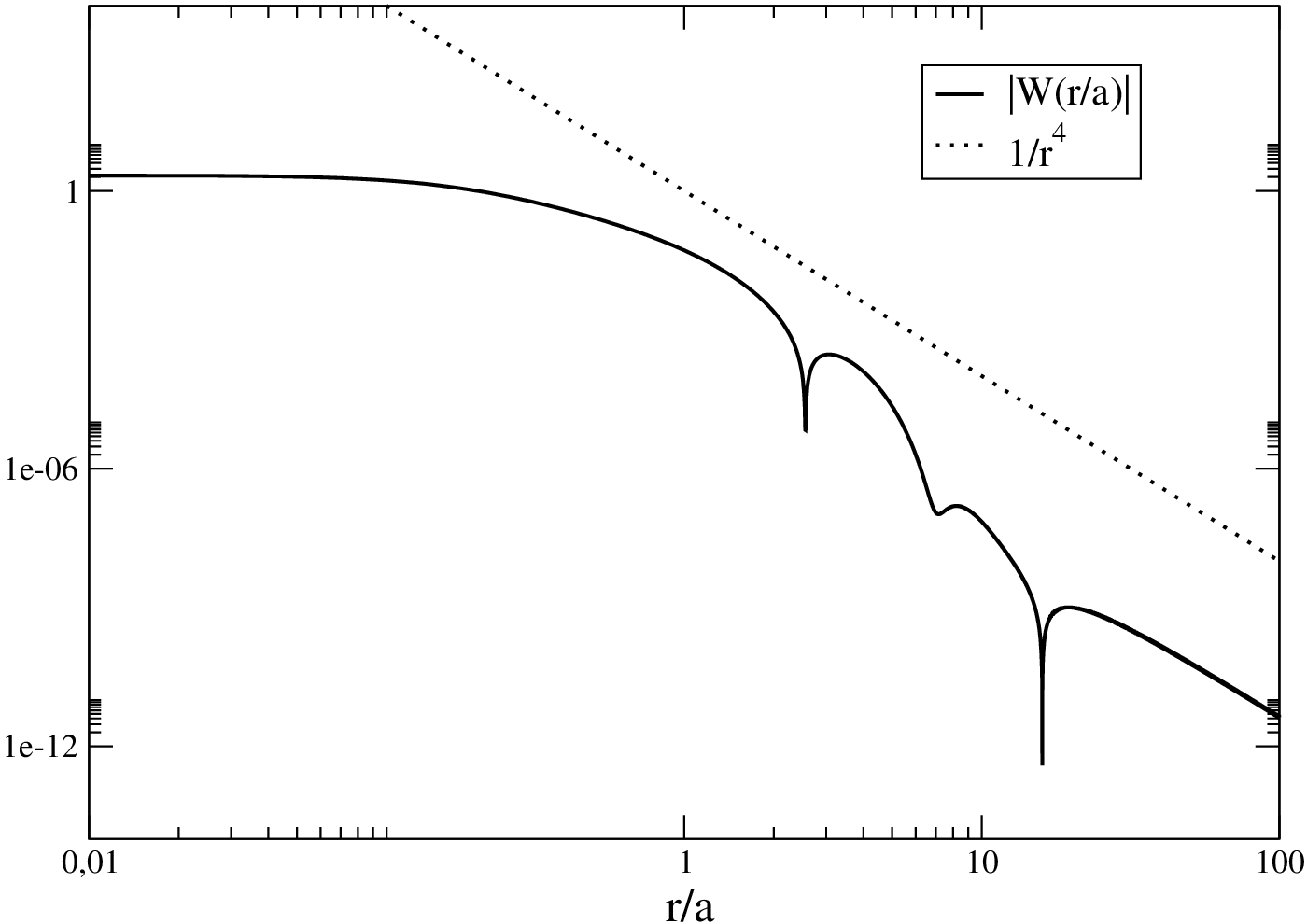}}
\caption{The window function in real space $\lvert W_{k_d, n_0}(r)\rvert$.\label{w}}
\end{figure}

As discussed in Sect.~\ref{Smoothing of discrete distributions}
the theoretical PS $P(k)$ can be interpreted as that of
a continuous mass distribution obtained by a physical 
smoothing with a function, whose Fourier transform is
given by 
\begin{equation}
\label{smoothing_HZ}
\lvert W_{k_d, n_0}(k)
\rvert^{-2} =1+\frac{\left(1+(A k)^\alpha\right)(1-e^{-k^2/2k_d^2})}
{n_0 {\cal N}k}\exp(k/k_c). 
\end{equation}
The corresponding real space smoothing function, 
calculated numerically for our chosen $S(k)$, is shown in Fig. \ref{w}.
It decays at large separation faster than $1/r^4$, and is
thus a localised smoothing in the sense we discussed
in Sect. \ref{discretisation}. It has, however, the
rather unsatisfactory feature of oscillating through 
negative values, albeit when the amplitude is already 
very small. We could, in principle, remedy this
by making a slightly different (but more complex)
choice of $D(k)$, but we do not anticipate that
it should cause any significant change in our results.

\begin{figure}
\resizebox{8cm}{!}{\includegraphics*{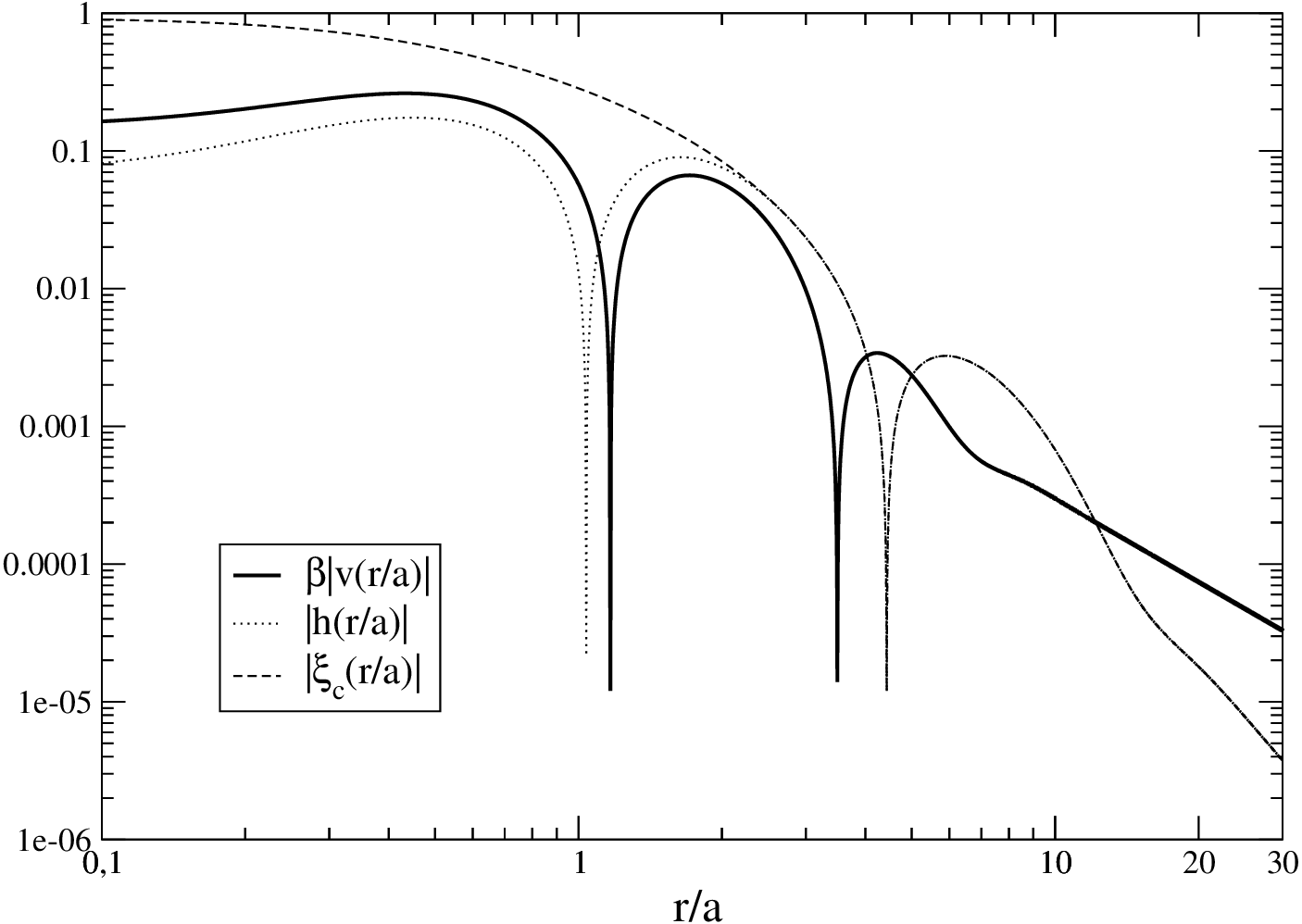}}
\caption{The correlation function, discretised correlation function,
and interaction potential obtained by the
inversion of the HNC for a PS as given in
Eq.(\ref{P_simpl}), with ${\cal N}=20\,a^4$, $A\approx 0.69\,a$ 
$k_c=2.7/a$ and $k_d=1.55/a$.\label{potential_toy}}
\end{figure}

Given our chosen $S(k)$ we can now use the 
inverted HNC equation Eq.~(\ref{inv_hnc}) to determine the 
combination $\beta v(r)$. The results are shown in  
Fig. \ref{potential_toy}. For $r/a>5$ the potential 
approaches the $1/r^2$ form corresponding to the 
small $k$ behaviour of the PS. 
For $r \gsim 4 a$ 
the correlation function is negative (i.e. the system is
anti-correlated) while it becomes positive
for smaller scales. Correspondingly 
the potential becomes attractive at approximately this scale.
At a smaller scale $r < lsim 2a$, we 
see that the continuous correlation function $\xi_c(r)$ (i.e. 
the Fourier transform
of Eq.~(\ref{P_simpl})) begins to deviate  
from the two-point correlation function $h(r)$ of the 
discretisation.

As anticipated, given the characteristics of $S(k)$ which 
describes a positively correlated system at small scales
(associated with the negative power law behaviour 
for $k > k_t$), the potential $v(r)$ is attractive at 
small scales.  Such a potential does not have a well defined 
thermal equilibrium \cite{ruelle}. To define
a  system whose thermal equilibrium gives the desired 
$S(k)$, we therefore add a repulsive core at some scale.
We take a core of the form $v_c(r)=0.2a^{10}/r^{12}$
(in units $Ze=1$) which has a negligible effect on
the potential at and above the scale $a$. We check that 
this full potential, used in the direct HNC, indeed 
reproduces the input $S(k)$ to a very 
good approximation. This is also a good
consistency check on the use of the HNC  for 
the chosen parameter values, in particular  
the normalisation given by the chosen value of
${\cal N}$. If ${\cal N}$ were too large we would
no longer be in the regime of weak correlations 
for which the HNC equation is valid.

We finally perform a simulation of molecular dynamics
with the determined potential to obtain configurations 
of points with the PS desired. The HNC equation 
gives the potential times the temperature $\beta v(r)$,
so we must choose a temperature. Any choice which keeps us in the
regime of validity of the HNC should be appropriate.
We make the simple choice $\beta=a^2$ (in units 
$Ze=1$ i.e. in which the large separation limit
of the potential is always simply $v(r)=1/r^2$)
which gives $\Gamma'=1$ in Eq.~(\ref{gamma-prime})
where, as we have discussed, the HNC gives very
accurate results for the $1/r^2$ potential
\footnote{Once we have length scales in the potential, as
we now do, the model (and in particular the conditions
of applicability of the HNC) is evidently no longer 
characterised  by the single parameter $\Gamma'$.
However we have here $|v(r) r^2| < 1$
down to the very small scales at which the hard-core 
becomes relevant i.e. the potentials can be considered 
as weaker than in the simple $1/r^2$ case. We
thus expect the HNC to work if it does for the
simple $1/r^2$ case at the same temperature. As always
here, however, it is the simulation of the molecular dynamics
which must ultimately validate the results obtained
with the HNC.}.

In the simulation we use this temperature 
to fix the initial velocities, taking them to
be random, and Gaussian distributed, with the 
corresponding variance.
Typically the system thermalises at a slightly
different temperature (as there is also 
a change in potential energy relative to
the initial configuration). By trial and 
error one can converge on the initial 
dispersion which gives the desired 
final temperature. For these simulations
this involved typically a few trials.
We will return to this point briefly
in the final section below.

\begin{figure}
\resizebox{8cm}{!}{\includegraphics*{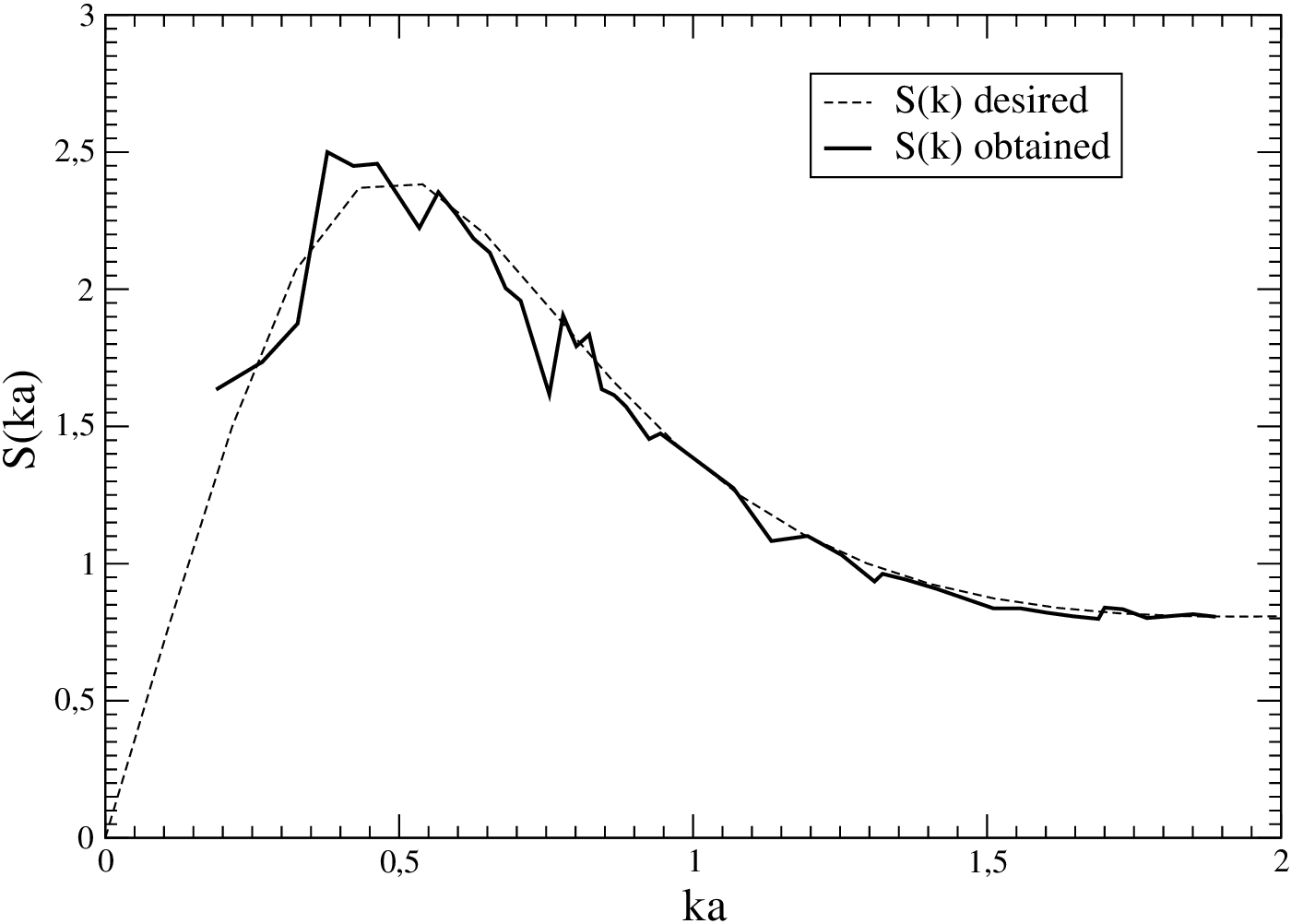}}
\caption{The PS  measured in a 
simulation of the molecular dynamics of $8788$
particles for the potential shown in the
previous figure. Also shown is the input
PS i.e. the PS of a system at equilibrium
with this potential as calculated in
the HNC.\label{CDM_P}}
\end{figure}

\begin{figure}
\resizebox{8cm}{!}{\includegraphics*{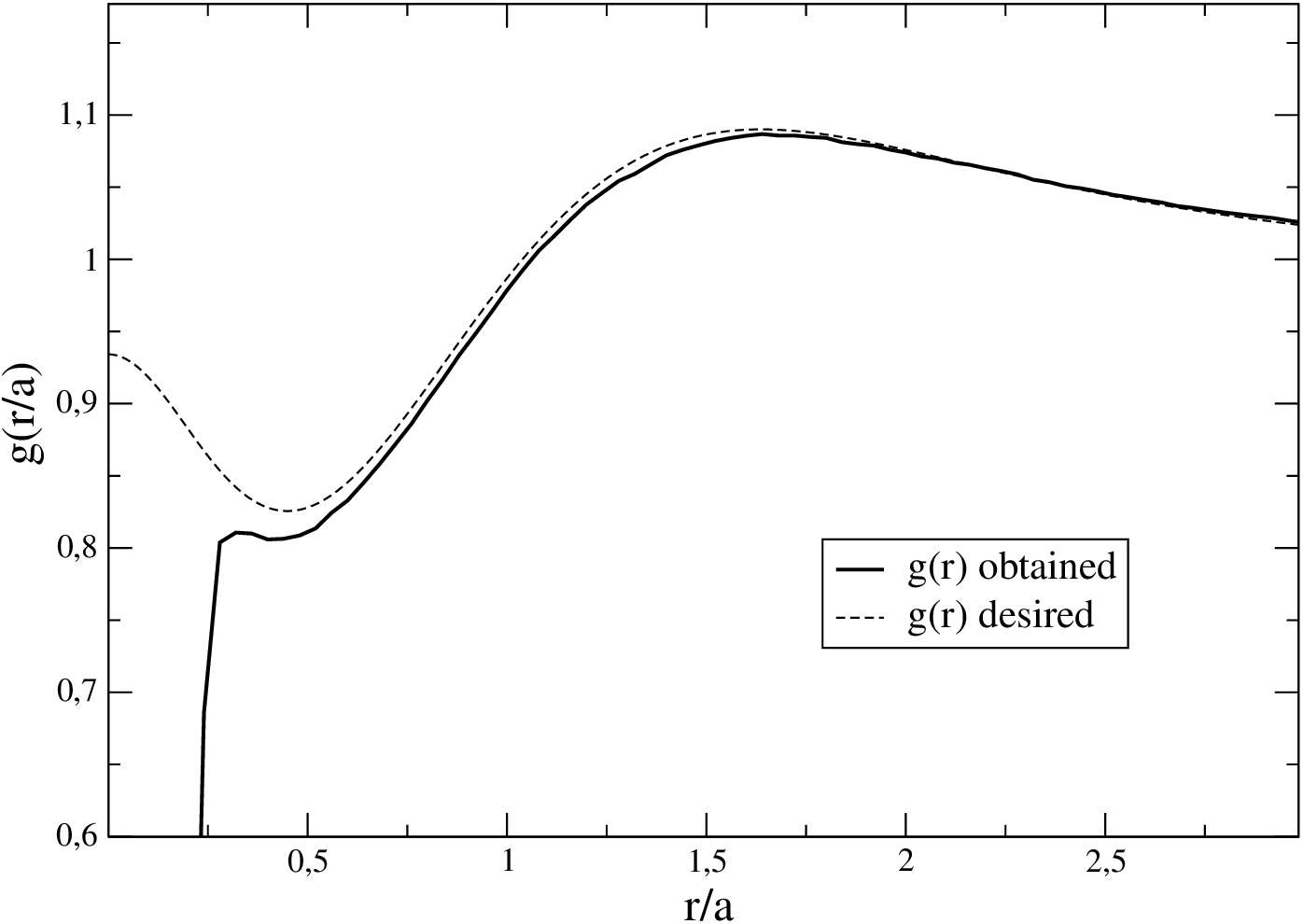}}
\caption{The real space correlation function
for the same cases as in the previous figure. The abrupt break 
to anti-correlation at about
$r/a \approx 0.25$ comes from the hard core introduced
to ensure stability of the system.\label{CDM_g}}
\end{figure}

In Fig. \ref{CDM_P} and Fig. \ref{CDM_g} are shown the
results for molecular dynamics simulations with
the potential determined from the HNC as
described above (and for a final temperature
s.t. $\beta=1/a^2$). 
The agreement is satisfactory, in the sense that we
reproduce quite well the correlation properties of the discretisation
we have defined, in both real and reciprocal space.
The effect of the hard-core is visible in real
space in the abrupt break in the correlation function,
and it  has a negligible effect above the scale 
$r \approx 0.2 a $ below which it dominates
\footnote{In fact we have found that, without 
the hard-core, our results are effectively unchanged 
at scales $r > 0.25 a$ and that the agreement with
the input correlation function extends to smaller scales.
Thus one infers that the time-scales of the instability to 
collapse must  be much  longer than those of 
the ``thermalisation'' of the system. This behaviour 
has been found also in simulations of attractive 
potentials discussed in \cite{krakoviack}.}.

\subsection{CDM-type spectra: realistic model}

\begin{figure}
\resizebox{8cm}{!}{\includegraphics*{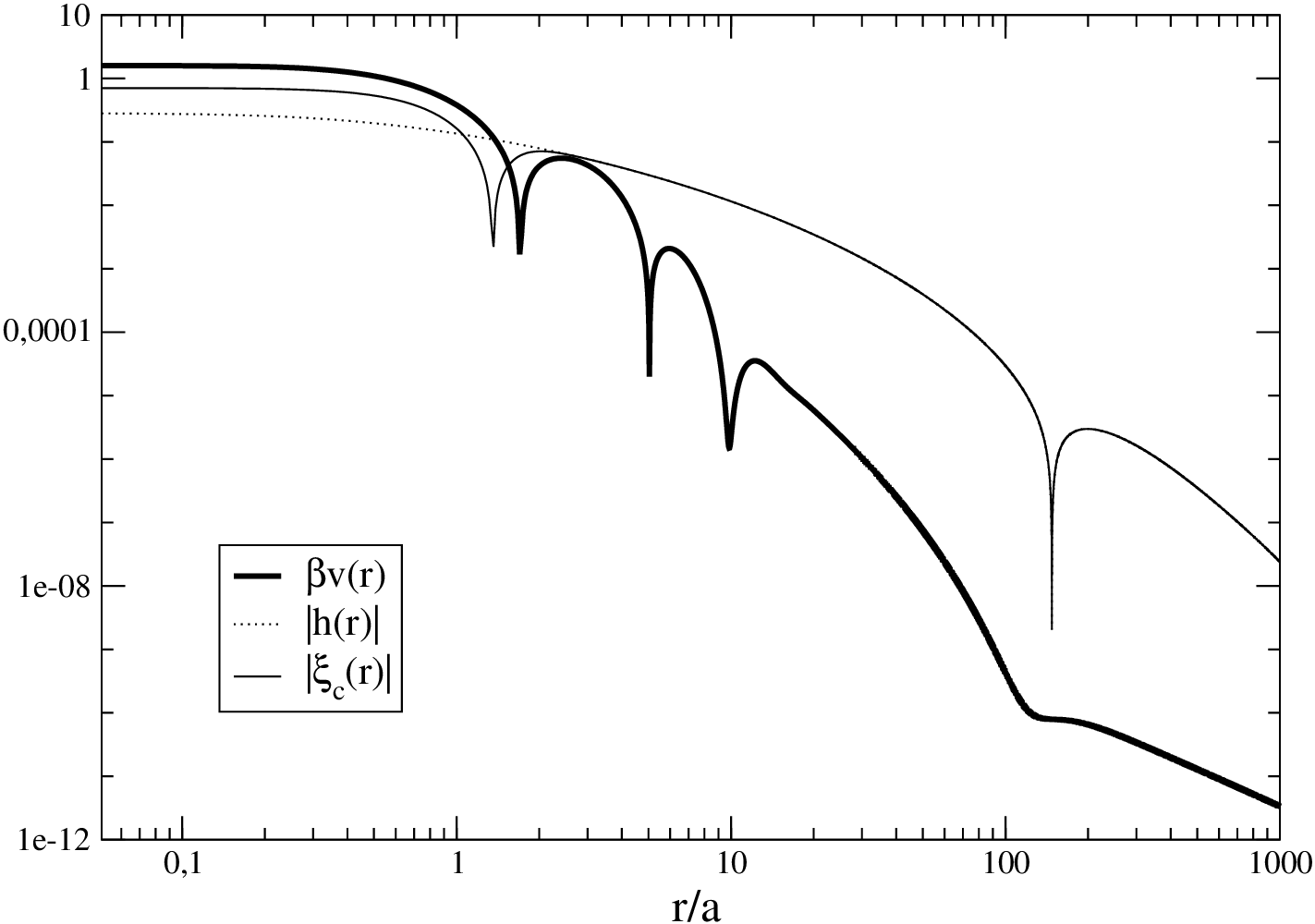}}
\caption{Correlation functions and interaction potential obtained for
the cosmological CDM spectrum described in the text.\label{potential_realistic}}
\end{figure}

We consider finally the determination of the potential which
should reproduce the cosmological PS (\ref{P_CDM}), with the parameters 
of a currently standard cosmological simulation (e.g. like
that taken as initial condition in the simulations of
the VIRGO consortium \cite{jenkins98}). To do so 
we must choose, in units of physical length,
the scale $a$ characterising the desired discretisation.
For our example, we choose this scale by supposing we have
the same physical density of point $n_0$ as in
some typical simulations of the VIRGO consortium:
we suppose that we have the particle density corresponding
to $256^3$ particles in a cubic box of side $239.5\mathrm{h^{-1}Mpc}$. 
The gives $a\approx 0.58 \mathrm{h^{-1}Mpc}$. We take our
initial time to correspond to red-shift $z=50$, and 
fix the normalization of the model at this time by 
the prescription that, using the
extrapolation of linear theory, one obtains today (at $z=0$)
$\sigma_8=1$.
This corresponds to a normalisation such that 
$\sigma_8=1/(z+1)=1/51$, and the normalization factor is then
${\cal{N}}=29381\mathrm{(h^{-1}Mpc)^4}$. 
The discretisation scale $k_d$ introduced is chosen again at the 
bounding value  $k_d\approx 1.55/a$.  In 
Fig. \ref{potential_realistic} are shown the 
correlation functions and the potential for this case. We note the 
same $1/r^2$ behaviour at large scales, but the potential is more 
complicated at small scales due to
the oscillations in the direct correlation function $c(r)$.
By simulating the molecular dynamics with this potential
with a sufficiently large number of particles, as we have done 
for a smaller number of particles for the simpler cases, we 
should  obtain a discretisation of the model with the 
properties Eqs.~(\ref{convergence1}) and (\ref{convergence2}).

\section{OTHER ISSUES AND CONCLUSIONS}

We have presented a new method to generate discrete distributions with
desired two-point correlation properties. It provides a promising 
alternative to the standard method for generating initial conditions 
for N-body simulation  in cosmology used, which involves displacing 
particles in a prescribed manner of a perfect lattice 
(or, sometimes, ``glassy'' configuration). As
discussed in detail in \cite{jm04} this latter method usually represents 
well the input theoretical PS in Fourier space at wave-numbers below the
Nyquist frequency, but produces in real space a system
with correlation properties which are a mixture of those of the
initial unperturbed lattice configuration and those of the theoretical
model. One obtains, in particular, a two-point correlation function
which is a rapidly oscillating function up to very large separations,
which is a behaviour completely different to that of the theoretical
model. In comparison the interest of this new method is
thus that it can give (by construction) a faithful representation of
the two-point statistical properties of a CDM-like spectrum in both
real and Fourier space. In particular, in the examples we have
considered, the correlation function in real space converges very well
to the theoretical input correlation function at a scale slightly
larger than $a$. Increasing the number of particles for the same 
physical size of
the system, the interparticle distance $a$, and thus this scale,
diminishes. In Fourier space the agreement is good (by construction)
for wavenumbers less than roughly a factor of two smaller than the
inverse of the scale $a$.

In conclusion we now discuss the remaining steps which are required 
to apply the method fully in the context of cosmological simulations. 
Specifically we address the question of the generation of initial 
velocities required for
such simulations, and then that of the modifications necessary 
to produce much larger distributions than those described here.  
Before doing so we first summarise briefly the method. 

\subsection{Summary of the method}

The principal steps involved in going from an initial input theoretical PS 
P(k) to an output configuration of initial conditions are the following
\footnote{About P(k) we assume only that it has the behaviour qualitatively
like that of almost all current cosmological models (i.e. a behaviour
$P(k) \sim k$ at small $k$ followed by a turn-over to a
monotonically decreasing behaviour.)}:

\begin{itemize}
\item{\bf 1.} {\it Determination of the function $S(k)$, the PS of the 
discretisation of $P(k)$ which we wish to generate}.  
As discussed in Sect.~\ref{discretisation} this choice is 
necessary as the PS of the point distribution cannot be {\it equal}
to $P(k)$, because continuous and discrete PS have intrinsically
different asymptotic properties at large $k$. The relation between
the two is given by Eq.~(\ref{prescription}), which,
requires the choice of a smoothing function $D(k)$. We have 
considered the simple choice $D(k)=1-e^{-k^2/2k_d^2}$, and
have taken $k_d$ as given by its upper bound as in 
Eq.~(\ref{value-k_d}).  The passage from $P(k)$ to its 
discretisation $S(k)$ involves therefore the  introduction
of only one parameter, a length scale which we have chosen
to take as $a$, determined by the particle density
$n_0$ through Eq.~(\ref{ion-sphere-radius}).

\item{\bf 2.} {\it Choice of parameter values in S(k)}.
Given the functional form of $S(k)$ one must give the
specific model by choosing the values of the parameters.
This means simply that the length scale $a$ must be associated
to a physical scale of the model (or, equivalently, that
the dimensionful parameters in $P(k)$ be specified
in units of $a$). In making this choice of parameters
there are two essential points. Firstly, the system will
be simulated with a finite number of particles
$N$, and thus a finite box (of side
$L=\left(\frac{4 \pi}{3} N \right)^{\frac{1}{3}} a $).
The parameters choices should evidently mean 
that the important characteristic scales in $P(k)$ fall
well inside the range $k_f < k < k_d$ (where $k_f$ is
the fundamental mode of the box). Secondly
the overall amplitude of $P(k)$ should be specified 
by a choice of the dimensionless parameter $n_0 P(k_d)$,
and should make this parameter of order unity.

\item{\bf 3.} {\it Determination of the two-body potential}.
Given a specific $S(k)$ the inversion procedure using the 
HNC, described in detail in Section~\ref{hnc}, is used to determine 
$\beta v(r)$ (where $v(r)$ is the two-body interaction potential).
If this associated potential does not give a stable thermal
equilibrium, a hard core is added well below the scale $a$ 
to ensure stability. Using this full potential (i.e. obtained 
by adding the core) one verifies that the direct HNC give
the desired $S(k)$. This is also a check on the applicability
of the HNC at the chosen amplitude.

\item{\bf 4.} {\it Generation of realizations of the discretisation}.
The molecular dynamics of $N$ particles interacting
with the potential $v(r)$ is simulated, starting from
initial velocities corresponding to a chosen temperature.
A simple appropriate choice of the latter is $\beta = 1/a^2$ 
(in units in which the long-range part of the potential
is $v(r)=1/r^2$). If the system thermalises at a slightly
different temperature, the initial velocities can be
appropriately modified by trial and error.   
  

\end{itemize}

\subsection{Generation of initial velocities}

In the initial conditions of cosmological simulations
of gravity, velocities must evidently also be ascribed 
to the particles of the distribution. We will now explain
that these velocites can be generated for our point distributions 
using the same method as in the standard algorithm.

In cosmological models small
initial density perturbations evolve, in linear theory, 
as a combination of a growing and a decaying mode.
The choice of velocities which is almost always 
appropriate corresponds to the selection of only
the growing mode. In practice this is implemented
using the Zeldovich approximation \cite{zel, buchert}, 
which gives the motion of a fluid element in a self-gravitating
fluid as 
\begin{equation}
\label{ZA}
{\vec x}({\vec q},t)={\vec q}  + f(t) {\vec u}({\vec q})
\end{equation}
where ${\vec x}$ is the (Eulerian) position of the 
fluid element, ${\vec q}$ its Lagrangian coordinate
and  $f(t)$ is a function of (cosmic) time $t$.
In an expanding universe ${\vec x}$ is
the comoving coordinate of the fluid element, related
to its physical coordinate ${\vec r}$
by ${\vec r}=a(t) {\vec x}$. 
We will suppose that ${\vec q}$ is the (comoving) 
position of the fluid element at a time $t_0$ 
i.e. $f(t_0)=0$. 

Using Eq.~(\ref{ZA}) in the continuity equation,
expanded  to linear order in the 
displacement field ${\vec u}$, gives
\begin{equation}
\rho(\vec{x},t)= \rho(\vec{q})
\left[1-f(t){\vec \nabla}_q \cdot{\vec u} \right]\,.
\label{continuity}
\end{equation}
where $\rho(\vec{x},t)$ and $\rho(\vec{q})$ are 
the mass densities in Eulerian and Lagrangian
coordinates respectively. The peculiar
gravitational field, defined by 
${\vec g}=\ddot{\vec r} - \frac{\ddot{a}}{a}{\vec r}$
obeys the Poisson-type equation 
\begin{equation}
{\vec{\nabla}}_r \cdot{\vec{g}}=-4\pi G \left(\rho({\vec r},t)-\frac{\rho_0}{a^3}\right)
\label{Poisson}
\end{equation}
sourced by the density fluctuations
in matter about the mean density ($\rho_0$ is the constant
mean comoving matter density of the universe, equal to the 
mean density in physical coordinates when $a=1$), and is 
given, using Eq.~(\ref{ZA}), by
\begin{equation}
\label{grav-ZA}
{\vec g}=a(t) [\ddot{f}(t) + 2 H(t) \dot{f} (t)] {\vec u}({\vec q})
\end{equation}
where $H(t)=\dot{a}/a$.
Substituting this relation in Eq.~(\ref{Poisson})
at $t=t_0$ (when $\vec{x}=\vec{q}$), and then in 
Eq.~(\ref{continuity}),  
it is easy to verify that, to linear order in the density 
perturbations about the mean comoving 
density $\rho_0$,
\begin{equation}
\frac{\rho(\vec{x},t)-\rho_0}{\rho_0}
=\left[ 1+\frac{4\pi G \rho_0 f(t)}
{\ddot{f}(t_0) + 2 H(t_0) \dot{f} (t_0)} \right] 
\frac{\rho(\vec{q})-\rho_0}{\rho_0}
\label{ZA-density}
\end{equation}
where we have chosen $a(t_0)=1$.

The evolution, on the other hand, of density fluctuations 
in Eulerian linear theory is given by 
\begin{equation}
\frac{\rho(\vec{x},t)-\rho_0}{\rho_0}
=D(t) \frac{\rho(\vec{q})-\rho_0}{\rho_0}
\label{linear-theory}
\end{equation}
where $D(t)$ is a solution of 
\begin{equation}
{\ddot{D}(t) + 2 H \dot{D} (t)} -\frac{4\pi G \rho_0}{a^3} D(t)=0\,,
\end{equation}
normalised to  $D(t_0)=1$. Thus taking 
$f(t)=D(t)-D(t_0)$ in Eq.~(\ref{ZA-density}),
one obtains Eq.~(\ref{linear-theory}) i.e. 
from Eq.~(\ref{ZA}) with $f(t)=D(t)-D(t_0)$  
one recovers Eulerian linear theory for
the evolution of small amplitude density fluctuations.
In particular choosing the growing mode solution
(i.e. $D_+(t)=(t/t_0)^{\frac{2}{3}}$ for a flat
matter dominated cosmology) gives fluctuations 
evolving in the pure growing mode as usually 
desired.

Using Eq.~(\ref{ZA}) the physical velocity 
of a fluid element is $\dot{\vec r} =H{\vec r} +
{\vec v}$, where 
\begin{equation}
\label{vel-ZA}
{\vec v}=a(t) \dot{f}(t) {\vec u}({\vec q}) 
\end{equation}
is its peculiar velocity. Now through 
Eq.~(\ref{grav-ZA}) it follows that
\begin{equation}
\label{vel-grav}
{\vec v}=\frac{\dot{f}(t)}{\ddot{f}(t) + 2 H(t) \dot{f} (t)} 
\,{\vec g}\,.
\end{equation}

To fix the initial velocities in our case we therefore
simply use the Zeldovich approximation as given by 
Eq.~(\ref{ZA}), the ${\vec q}$ being identified with the 
initial positions of our particles in the initial 
configuration corresponding to the required initial
time $t_0$.  Eq.~(\ref{vel-grav}) evaluated at $t=t_0$
and with $f(t)=D_+(t)$ specifies the velocities,
in terms of the peculiar gravitational field ${\vec g}$ 
at each point in the initial configuration. 

It is perhaps useful to clarify the relation of this
use of the Zeldovich approximation to the usual one
in setting up initial conditions for cosmological
NBS. To do so we write Eq.~(\ref{ZA}) as 
\begin{equation}
\label{ZA-variable-change}
{\vec x}({\vec R},t)={\vec R}  + D_+(t) {\vec u}({\vec R})
\end{equation}
by a change of the Lagrangian variable 
${\vec R}={\vec q} - {\vec u}({\vec q})$
\cite{buchert}.
One has $D_+(t \rightarrow 0)=0$
so that the  density in the coordinates ${\vec R}$ is that 
of the universe as $t \rightarrow 0$. 
In the standard method of setting up initial conditions
in cosmological simulations, these initial
positions ${\vec R}$ are taken to be those of points 
in a ``uniform'' configuration: usually a lattice, or
sometimes also a ``glassy'' configuration \cite{white}.
The initial conditions (at $t=t_0$) are then generated by
moving the points to 
${\vec q}={\vec R}+{\vec u}({\vec R})$
with the displacements ${\vec u}({\vec R})$ given
through Eq.~(\ref{grav-ZA}), the gravitational field 
being evaluated at the position ${\vec R}$ by a realization 
of the desired gravitational potential field (which in turn 
is determined by the theoretical power spectrum 
through the Poisson equation). This is accurate
to linear order in the displacement field.

The crucial point which allows us to use the Zeldovich approximation 
is that this approximation does not require, as is supposed in its 
usage just described in the standard algorithm for generating 
initial conditions, that the evolution of the fluid start 
from a perfectly homogeneous configuration. Indeed the 
Zeldovich approximation is simply an asymptotic solution
of the evolution of a self-gravitating fluid away from
a generic initial configuration with some initial velocity
and gravitational field \cite{buchert}. More specifically 
all that is required to recover the linear amplification of
density fluctuations which follow from the Zeldovich approximation, 
as described in the first part of this section,  is that the
density fluctuations are small. In this respect, we note that one can 
actually  consider the standard algorithm as a two step process 
in complete analogy to our method. In the first step particles are displaced
off a lattice (or other ``uniform'' particle  configuration) 
in order to produce a distribution with a power
spectrum resembling as closely as possible
that of the theoretical power spectrum $P(k)$.
This does not in fact require the use of the Zeldovich
approximation, but only the continuity  equation. 
Indeed from Eq.~(\ref{ZA-variable-change})
with $D(t_0)=1$, and assuming $\rho(\vec{R})=\rho_0$,
the continuity equation gives
\begin{equation}
\frac{\rho(\vec{x},t)-\rho_0}{\rho_0}
=-{\vec {\nabla}}_R  \cdot {\vec u}({\vec R})\,.
\label{ZA-density-standard}
\end{equation}
From this it follows that it is sufficient to 
take ${\vec u}({\vec R})$ to be 
the gradient of a scalar field $\phi(\vec{R})$ (as 
any rotational component for ${\vec u}$ will
not generate density fluctuations). Through
Eq.~(\ref{ZA-density-standard}) we have
that 
$|\tilde{\phi}(\vec{k})|^2=P(k)/k^4$, and thus
that $\phi(\vec{R})$ is just proportional to the
gravitational potential. The procedure is thus
precisely the standard one. The second step then 
consists in generating the
velocities, and one can proceed exactly as we
have described for our method, using now
explicitly the Zeldovich approximation and,
specifically, Eq.~(\ref{vel-grav})
with the gravitational field ${\vec g}$ given
at each point in the perturbed configuration
(i.e. at the Lagrangian coordinate $\vec{q}$).
To the linear order in the displacements this
is identical to the standard prescription.

\subsection{Generation of larger distributions}

To test the proposed method here we have generated
particle distributions with at most of order ten
thousand particles, a number typical in statistical
physics for the simulation of the molecular dynamics of
systems of the type we have used (and for which the
numerical codes have been conceived).
It may be interesting and useful to evolve
configurations of this size generated in this way,
and by the standard method, under their self-gravity
and to do a comparative study of this evolution. 
However, to use it as a truly alternative method
in full cosmological simulations, it will be necessary
to generate much larger configurations: in this context 
simulations typically now involve, at a minimum, hundreds 
of thousands of particles. 

The Ewald summation method we have used here 
can be shown \cite{perram-scaling-ewald} to scale 
(for a fixed precision on the calculation of the energy and forces)
as $N^{3/2}$. Given that the largest simulations we have reported here,
of approximately nine thousand particles, ran in about one day 
on a typical current PC, we could simulate only a slightly larger 
number in a reasonable time with our code. There are, however, 
well documented ways of speeding up the Ewald sum method by 
calculating the reciprocal space part of the sum on a grid. 
One can then employ rapid FFT algorithms, which scale 
as $N \log N$ \cite{Ewald-FFT}, with the direct space part of the
sum scaling as $N$. There are numerous 
variants (e.g. P3M, PME, SPME), which differ 
primarily in how the charges (or masses) are
assigned to the grid. 
Going beyond the Ewald method, or in combination with it,
an even faster algorithm is the Fast Multipole
Method (FMM) (see e.g. \cite{multi-pole methods}).  It is
based on a decomposition of the system into sub-volumes, with the
interaction between the charges contained in a given volume
with those in a sufficiently distant sub-volume being calculated 
with a multipole expansion.  Most of these techniques are of
course known in the context of cosmological NBS
(e.g. \cite{EDFW}, \cite{couchman}) where they have 
been applied to gravity. Note, however, that their use if not
limited to this case, and in particular does not require 
that the potential necessarily obey a local equation like the
Poisson equation as for the $1/r$ potential.

\subsection{Other remarks}

Another possible improvement concerns the 
choice of initial velocities in our simulations
of the molecular dynamics.
When introducing the potential calculated with HNC we 
implictly choose the equilibrium temperature before performing
the MD simulation. Thus, as explained above, we put the initial 
velocities to get the chosen final temperature. The problem is 
that we do not know a priori how the system is going to reach 
equilibrium and it is necessary to do trials with different
initial velocities until the desired equilibrium temperature
is attained. A solution to this problem would be to modify the MD 
to work in the canonical ensemble (in which the temperature is
fixed) rather than in the micro-canonical ensemble as we 
have done here.

We remark finally that we have not attempted here to address at all 
the very interesting underlying question motivating this work
of the importance of the accuracy and nature of 
the discrete  representations of cosmological IC in the dynamical evolution
under gravity of these IC. Our purpose  here has been 
to provide an alternative method for representing IC
which should allow such questions to be 
addressed with a new method, and hopefully answered, in 
due course.

We thank J.M. Caillol, A. Gabrielli and F. Sylos Labini 
for useful discussions and comments.

\end{document}